\documentclass[runningheads]{llncs}
\usepackage{amsmath}
\usepackage[utf8]{inputenc}
\usepackage{authblk}
\usepackage{amssymb}
\usepackage{enumitem}
\usepackage{booktabs}
\usepackage[show]{chato-notes}
\usepackage{graphicx}
\usepackage{verbatim}
\usepackage{hyperref}
\usepackage{mathtools}
\usepackage{multirow}
\usepackage{tikz}
\usepackage{pgfplotstable} 
\graphicspath{{plots/}}
\usepackage{subfig}
\usepackage{float}
\usepackage{amsmath} 
\usepackage[utf8]{inputenc} 
\usepackage{hyperref} 
\usepackage{graphicx} 
\usepackage{listings} 
\usepackage{circuitikz}
\usepackage{hyperref}
\usepackage[linesnumbered,ruled]{algorithm2e}
\usepackage{multicol}
\usepackage{etoolbox}
\patchcmd{\paragraph}{\itshape}{\bfseries\boldmath}{}{}
\usepackage{csvsimple}
\usepackage{siunitx} 
\usepackage{pgfplotstable} 

\usepackage{textcomp}
\usetikzlibrary{shapes,arrows}

\usepackage{amssymb}

\usetikzlibrary{fit,positioning,arrows,automata,calc}
\tikzset{
  main/.style={circle, minimum size = 5mm, thick, draw =black!80, node distance = 10mm},
  connect/.style={-latex, thick},
  box/.style={rectangle, draw=black!100}
}

\title{Deep unfolding for non-negative matrix factorization with application to mutational signature analysis}
\author{Rami Nasser\inst{1} \and
Yonina C. Eldar\inst{2} \and
Roded Sharan\inst{3}$^,$\thanks{Corresponding author. Email: \url{roded@tauex.tau.ac.il.}}}
\institute{
Dept. of Statistics and Operations Research, Tel Aviv University, Tel Aviv 69978, Israel \and
Department of Math and Computer Science, Weizmann Institute of Science, Rehovot, Israel \and
Blavatnik School of Computer Science, Tel Aviv University, Tel Aviv 69978, Israel}

\authorrunning{R. Nasser et al.}
\titlerunning{Deep unfolding for non-negative matrix factorization}

\begin{document}
\maketitle
\begin{abstract}
Non-negative matrix factorization (NMF) is a fundamental matrix decomposition technique that is used primarily for dimensionality reduction and is increasing in popularity in the biological domain. Although finding a unique NMF is generally not possible, there are various iterative algorithms for NMF optimization that converge to locally optimal solutions. Such techniques can also serve as a starting point for deep learning methods that unroll the algorithmic iterations into layers of a deep network. Here we develop unfolded deep networks for NMF and several regularized variants in both a supervised and an unsupervised setting. We apply our method to various mutation data sets to reconstruct their underlying mutational signatures and their 
exposures. We demonstrate the increased accuracy of our approach over standard formulations in analyzing simulated and real mutation data.
\end{abstract}

\section{Introduction}
Non-negative matrix factorization (NMF) is a popular and useful decomposition tool for high dimensional data. It is widely used in signal and image processing, text analysis and in analyzing DNA mutation data. NMF is NP-hard~\cite{vavasis2010complexity} in general, and is commonly approximated by various iterative algorithms such as multiplicative updates~\cite{Lee} and ANLS~\cite{Lin2007-tf}. Almost all NMF methods use a two-block coordinate descent scheme, which alternatively optimizes one of the $W,H$ matrices in the data decomposition $V\sim WH$ while keeping the other fixed~\cite{gillis2014and}. These iterative algorithms generally suffer from slow convergence and high computational cost when applied to large matrices~\cite{Kim2011-fj}.  

Recently, architectures based on deep learning were suggested for NMF~\cite{Hershey2014-xw,Wisdom2017-ce} as part of a general unfolding (or unrolling) framework~\cite{monga2019algorithm}. Unrolling techniques connect between iterative methods and deep networks by viewing each iteration of an underlying iterative algorithm as a layer
of a network, such that concatenating the layers forms a deep neural
network where the algorithm parameters transfer to the network parameters.
The network is trained using back-propagation, resulting
in model parameters that are learned from real world training sets. However, these previous unrolling methods for NMF were limited to supervised settings where one of the matrix factors is known and can be used for training. 

Here we develop a deep unrolled network architecture, which we call DNMF, for regularized variants of NMF for both the supervised and unsupervised settings. In our model, we learn two types of weight matrices for added flexibility in learning complex patterns and design the network so that conventional back propagation tools such as the auto gradient in Pytorch can be used to allow for large scale implementation. We implement the resulting networks and show their utility over standard iterative formulations. 
In particular, we apply our constructions to analyze a diverse collection of simulated and real mutation data sets, and show that they lead to better reconstructions of unseen data compared to 
the multiplicative update scheme. In the supervised setting, we train the network based on given input vectors $V$ and their corresponding coefficients $H$, without the need of knowing the underlying dictionary (corresponding to mutational signatures) $W$. In the unsupervised setting, our network operates with the input non-negative data matrix $V$ only.

Our contribution is three-fold: (i) we develop a deep unfolded network formulation for regularized NMF; (ii) we generalize this formulation to support an unsupervised setting, in which NMF is typically applied; (iii) we show the network's utility over standard formulations in analyzing simulated and real mutation data sets.

\section{Methods}
\subsection{Problem formulation and current approaches}
NMF receives as input a non-negative matrix $V_{f \times n}$ and a number $k$ of desired factors; its goal is to decompose $V$ into a product of two non-negative matrices $W_{f\times k}$ and $H_{k\times n}$ such that
$\| V-WH \|_2$ is minimized.

A popular iterative method to approximate the above is Lee-Seung's multiplicative update (MU) scheme~\cite{Lee}:
    \begin{equation}
        H_{l+1} \gets H_l \odot \dfrac{W_l^TV }{W_l^TW_lH_l}
        \label{update h}
    \end{equation}
    \begin{equation}
         W_{l+1} \gets W_l \odot \dfrac{VH_l^T}{W_lH_lH_l^T}
        \label{update w}
    \end{equation}
where $\odot, \dfrac{[.]}{[.]}$ denote entry-wise multiplication and division, superscript $T$ denotes matrix transpose, and the subscript index denotes the iteration number. Usually, $W_0, H_0$ are initialized by random or fixed non-negative values; more complicated initialization strategies have also been introduced~\cite{Albright2006-wt,Boutsidis2008-vz}.

\paragraph{Regularized variants.}
Hoyer et al.~\cite{hoyer2002non} extended the classical multiplicative update scheme for the case of an $L_1$ penalty imposed on the coefficients of $H$. Other works have also developed formulations for $L_2$ regularization~\cite{Wang2016-ay}. For completeness, we redevelop a regularized variant
with both penalties in Appendix~\ref{proof}. Fixing $W$ and looking at one sample $v$ and one column $h$
of $H$ at a time, we consider the problem:
\begin{equation}
    \min_{h\geq{0}} \left\{\frac{1}{2}\| v - Wh \|_2 + \lambda_1\| h \|_1 + \frac{1}{2}\lambda_2\| h \|_2^2\right\}.
    \label{cost_fun}
\end{equation}
This leads to the following multiplicative update equation (see Appendix~\ref{proof}):
\begin{equation}
    h_{l+1} \gets h_l \odot \dfrac{W^Tv}{W^TWh_l+\lambda_1+\lambda_2h_l}.
\label{eq:update rule}
\end{equation}
Note that if $h_0,W,v$ and the regularization paramters $\lambda_1,\lambda_2$ are non-negative, then $h_l$ will be non-negative as well.

\subsection{Unrolling the iterative algorithm}
To obtain our suggested unrolled network, it will be convenient to consider one input sample $v\in{R^{f}}$ at a time. Following~\cite{Hershey2014-xw}, we develop the network architecture by optimizing the corresponding column $h$ while allowing $W$ to be part of the network's parameters that are being learned and, moreover, vary between layers. In the unrolled network, each layer represents a possible solution to $h$ that is formed by 
a non-linear transformation of the values at the previous layer. The transformation imitates the multiplicative update formula~\ref{eq:update rule}, with $W$ varying between the layers (rather than being fixed) and $\lambda_1,\lambda_2$ fixed across layers. Moreover, the network ignores the dependency between the $W^T$ term and the $W^TW$ terms in the update formula and treats them as independent matrices, $A$ and $B$, respectively. These matrices are later learned from data.
Overall, in the supervised setting, the network relies on training data $v_1, v_2, ... v_N \in R^f$ and their corresponding coefficient vectors $h'_1, h'_2, ... h'_N \in R^k$ to optimize the parameters $A_l, B_l, \lambda_1, \lambda_2$. The resulting network model is depicted in Figure~\ref{fig:Smodel}.

\tikzset{%
  block/.style    = {draw, thick, rectangle, minimum height =2em,
    minimum width = 2em},
  sum/.style      = {draw, circle, node distance = 2cm}, 
  input/.style    = {coordinate}, 
  output/.style   = {coordinate} 
}
\begin{figure}
\begin{tikzpicture}[auto, node distance=2cm]
\draw
    node at (-5.5,3.5)[circle, name=input1] {$h_l$}
    node [sum, right of=input1] (f) {$f$}
    node [circle, name=input2, below of=f] {$v$}
    node [circle, name=output1, right of=f] {$h_{l+1}$};
	\draw[->](input1) -- node {$B_l$}(f);
	\draw[->](input2) -- node {$A_l$}(f);
	\draw[->](f) -- node {}(output1);
	
	\matrix at (0,-1) (m1) [row sep=2.5mm, column sep=5mm]
	{
		\node[block]                       (m00) {$h_0$};     &
		\node[coordinate]                  (m01) {};          &
		\node[sum]                         (m02) {$f$};       &
		\node[coordinate]                  (ms1) {};          &
		\node[block]                       (m03) {$h_1$};     &
		\node[coordinate]                  (ms2) {};          &
		\node[sum]                         (m04) {$f$};       &
		\node[coordinate]                  (ms3) {};          &
		\node[block]                       (m05) {$h_2$};     &
		\node[circle]                      (m06) {...};       &
		\node[block]                       (m07) {$h_l$};     &
		                                                      \\
		\node[coordinate]                  (m10) {};          &
		\node[coordinate]                  (m11) {};          &
		\node[coordinate]                  (m12) {};          &
		\node[coordinate]                  (ms11) {};          &
		\node[coordinate]                  (m13) {};          &
		\node[coordinate]                  (ms12) {};          &
		\node[coordinate]                  (m14) {};          &
		\node[coordinate]                  (ms13) {};          &
		\node[coordinate]                  (m15) {};          &
		\node[coordinate]                  (m16) {};          &
		\node[coordinate]                  (m17) {};          &
                                                              \\
		\\
		\node[block]                       (m20) {$v$};       &
		\node[coordinate]                  (m21) {};          &
		\node[coordinate]                  (m22) {};          &
		\node[coordinate]                  (ms21) {};          &
		\node[coordinate]                  (m23) {};          &
		\node[coordinate]                  (ms22) {};          &
		\node[coordinate]                  (m24) {};          &
		\node[coordinate]                  (ms23) {};          &
		\node[coordinate]                  (m25) {};          &
		\node[circle]                      (m26) {...};       &
		\node[coordinate]                  (m27) {};          &
                                                              \\
	};

	\draw[->](m00) -- node {$B_0$}(m02);
	\draw[->](m02) -- node {}(m03);
	\draw[->](m03) -- node {$B_1$}(m04);
	\draw[->](m04) -- node {}(m05);
	\draw[->](m05) -- node {}(m06);
	\draw[->](m06) -- node {}(m07);
	\draw[->](m20) -- node {}(m26);
	\draw[->](m22) -- node {$A_0$}(m02);
	\draw[->](m24) -- node {$A_1$}(m04);
	
	\draw [color=gray,thick](-6,4.5) rectangle (-1,1);
	\node at (-5,5) [above=5mm, right=0mm] {\textsc{Single Layer}};
	\node at (2.5,3.5) [] {\textsc {$f =  h_l \odot \dfrac{A_lv}{B_lh_l+\lambda_1+\lambda_2h_l}$}};
\end{tikzpicture}
\caption{A sketch of the proposed supervised unrolled network for NMF.}
\label{fig:Smodel}
\end{figure}

To test the resulting network, we used $10$ layers (see Results for performance across varying depth values) and implemented back propogation using Pytorch. Training was performed through minimizing the MSE loss function $\|h_{10}-h'\|_2$ using the ADAM optimizer with learning rate $0.001$. The parameters were updated using constrained gradient descent to guarantee that network weights are non-negative. 

The model parameters including the $A,B$ matrices across layers and the two regularization parmeters $\lambda_1, \lambda_2$ were initialized to a fixed positive value (value of 1). We also initialized the entries of $h_0$ to the same value. For each of the data sets we trained a model based on $80\%$ of the data, and measured the MSE with respect to the remaining $20\%$ using the true matrix $H$.

\subsection{An unsupervised variant}
Typically, we do not know the decomposition matrices $H$ and/or $W$ in advance, in which case 
supervised training is not feasible. Instead, we propose to evaluate a solution by its
ability to reconstruct the original matrix $V$. To this end, after obtaining the network output $h$
for each of the data columns, we use
non negative least squares (NNLS) to reconstruct $W$~\cite{lawson1995solving} and adjust the cost function accordingly.
In detail, we start by initializing $h_0$ to fixed values for every column of $V$, the two columns are forward propagated in the network,  and the resulting $h_{\ell}$-s for all samples are gathered to form the estimated $H$ matrix. Next, we apply NNLS to estimate $W$ from $V$ and $H$. Last, we calculate the cost function
given in Equation~\ref{cost_fun} and backpropagate to update the network weights $A, B$.
The model is depicted in Figure~\ref{fig:um}. 
 
In the unsupervised case we cannot learn the regularization parameters as they affect the cost function
and if we would omit them from the cost function, their optimal value will be zero (corresponding to no regularization). Hence, 
in this variant we use $\lambda_1=\lambda_2=\lambda$ and present results for $\lambda=0,1,2$.

\tikzset{%
  block/.style    = {draw, thick, rectangle, minimum height =2em,
    minimum width = 2em},
  sum/.style      = {draw, circle, node distance = 2cm}, 
  input/.style    = {coordinate}, 
  output/.style   = {coordinate} 
}
\begin{figure}
\begin{tikzpicture}[auto, node distance=2cm]
\draw
    node at (-5.5,3.5)[circle, name=input1] {$h_l$}
    node [sum, right of=input1] (f) {$f$}
    node [circle, name=input2, below of=f] {$v$}
    node [circle, name=output1, right of=f] {$h_{l+1}$};
	\draw[->](input1) -- node {$B_l$}(f);
	\draw[->](input2) -- node {$A_l$}(f);
	\draw[->](f) -- node {}(output1);
	
	\matrix at (0,-1) (m1) [row sep=2.5mm, column sep=5mm]
	{
		\node[block]                       (m00) {$h_0$};     &
		\node[coordinate]                  (m01) {};          &
		\node[sum]                         (m02) {$f$};       &
		\node[coordinate]                  (ms1) {};          &
		\node[block]                       (m03) {$h_1$};     &
		\node[coordinate]                  (ms2) {};          &
		\node[sum]                         (m04) {$f$};       &
		\node[coordinate]                  (ms3) {};          &
		\node[block]                       (m05) {$h_2$};     &
		\node[circle]                      (m06) {...};       &
		\node[block]                       (m07) {$h_l$};     &
		\node[block]                       (m08) {$h$};       &
		                                                      \\
		\node[coordinate]                  (m10) {};          &
		\node[coordinate]                  (m11) {};          &
		\node[coordinate]                  (m12) {};          &
		\node[coordinate]                  (ms11) {};          &
		\node[coordinate]                  (m13) {};          &
		\node[coordinate]                  (ms12) {};          &
		\node[coordinate]                  (m14) {};          &
		\node[coordinate]                  (ms13) {};          &
		\node[coordinate]                  (m15) {};          &
		\node[coordinate]                  (m16) {};          &
		\node[sum]                         (m17) {NNLS};    &
		\node[block]                       (m18) {$W$};       &
                                                              \\
		\\
		\node[block]                       (m20) {$v$};       &
		\node[coordinate]                  (m21) {};          &
		\node[coordinate]                  (m22) {};          &
		\node[coordinate]                  (ms21) {};          &
		\node[coordinate]                  (m23) {};          &
		\node[coordinate]                  (ms22) {};          &
		\node[coordinate]                  (m24) {};          &
		\node[coordinate]                  (ms23) {};          &
		\node[coordinate]                  (m25) {};          &
		\node[circle]                      (m26) {...};       &
		\node[coordinate]                  (m27) {};          &
                                                              \\
	};

	\draw[->](m00) -- node {$B_0$}(m02);
	\draw[->](m02) -- node {}(m03);
	\draw[->](m03) -- node {$B_1$}(m04);
	\draw[->](m04) -- node {}(m05);
	\draw[->](m05) -- node {}(m06);
	\draw[->](m06) -- node {}(m07);
	\draw[->](m20) -- node {}(m26);
	\draw[->](m22) -- node {$A_0$}(m02);
	\draw[->](m24) -- node {$A_1$}(m04);
	\draw[->](m07) -- node {}(m17);
	\draw[->](m27) -- node {}(m17);
	\draw[->](m26) -- node {}(m27);
    \draw[->](m17) -- node {}(m18);
    \draw[->](m07) -- node {}(m08);
	
	\draw [color=gray,thick](-6,4) rectangle (-1,1);
	\node at (-5,5) [above=5mm, right=0mm] {\textsc{Single Layer}};
	\node at (2.5,3.5) [] {\textsc {$f = h_l \odot \dfrac{A_lv}{B_lh_l+\lambda_1+\lambda_2h_l}$}};
\end{tikzpicture}
\caption{A sketch of the unrolled deep network for the unsupervised variant.}
\label{fig:um}
\end{figure}
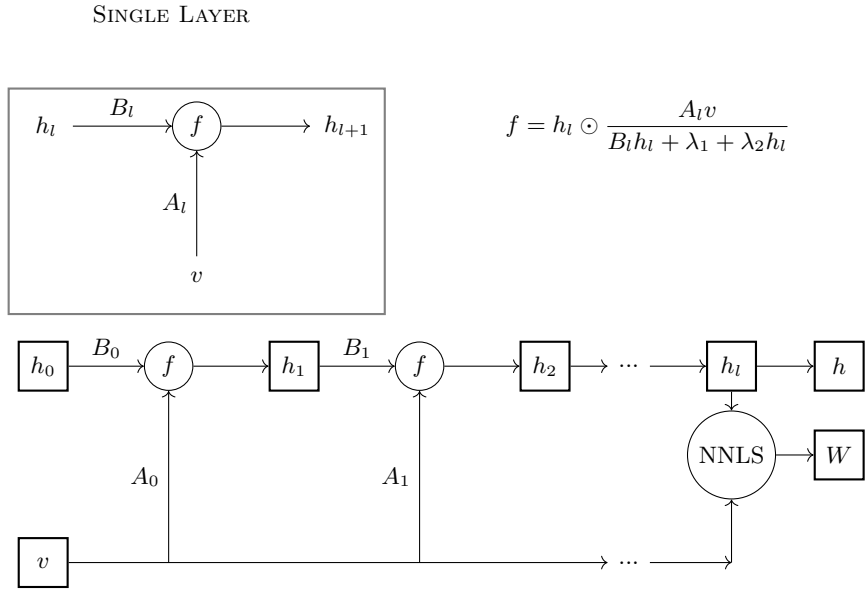

\subsection{Data description and performance evaluation}
We used two types of mutation data sets: simulated and real ones. In all cases the number of rows in the observed (count) matrix $V$ was 96, representing the 96 standard mutation categories~\cite{Alexandrov2019-fe}. For such data, $V$ is assumed to be the result of the activity of certain mutational processes whose signatures are given by the dictionary $W$ and whose exposures are given by the coefficient matrix $H$. We describe these data sets below.


\paragraph{Simulated data.}
The simulated data were taken from ~\cite{Alexandrov2019-fe} and includes multiple 
mutation data sets with varying numbers of underlying signatures and degrees of noise. For each data set we are
given an observed matrix of mutation counts (denoted $V$ above) and its decomposition into signature ($W$) and
exposure ($H$) matrices. In total, we used 12 different simulated data sets with at least 1,000 samples each as detailed in Appendix~\ref{appendix:data}.

\paragraph{Real data.}
We analyzed a breast cancer (BRCA) mutation  data set of whole-genome sequences from the International Cancer Genome Consortium (ICGC). The data set has 560 samples and believed to be the result of the 
activity of 12 mutational processes as cataloged in the COSMIC database.

\paragraph{Performance evaluation.}
We evaluated our method on each dataset using 5-fold cross-validation and compared to the standard multiplicative update method under various regularization schemes. All model parameters in both methods were initialized to one, unless specified otherwise.
In the supervised case, we report the MSE between the true $H$ and the estimated matrix $H_l$ over a test set (20\% of the samples), where the MSE is averaged over the columns of $H$. 
In the unsupervised case, we report the MSE between $V$ and its reconstruction $WH$ over a test set (20\% of the samples), where the MSE is averaged over the columns of $H$. For both DNMF and MU, $W$ in inferred using the training samples and $H$ is estimated on the test samples. For DNMF, the estimation of $H$ is done by propagating the columns of $V$ in the learned network. For MU, it is done by fixing $W$ and using the iterative update rule to compute $H$. 

\subsection{Implementation and runtime details}
All reported runs were done in Google Colab using a 2-core CPU (\verb=x86_64=).
Code is available at \url{https://github.com/raminass/deep-NMF}.
The inference time of supervised/unsupervised DNMF with 10 layers was 0.0019 sec, similar to 
a 10-iteration MU inference in the supervised case (0.0021 sec), and an order of magnitude faster
than a 100-iteration MU inference in the unsupervised case (as used here, 0.016 sec).

\section{Results}
We developed a deep learning based framework for non-negative matrix factorization, which we call DNMF. The DNMF framework imitates the classical multiplicative update (MU) scheme for the problem by unrolling its iterations as layers in a deep network model.
We further developed regularized variants for MU and DNMF. We apply our framework in both a supervised setting, where training data regarding the true factorization is available, 
and in an unsupervised setting. Full details on the different models appear in the Methods.

\begin{figure}
    \centering
    \subfloat [\centering  ]{\includegraphics[width=0.33\textwidth]{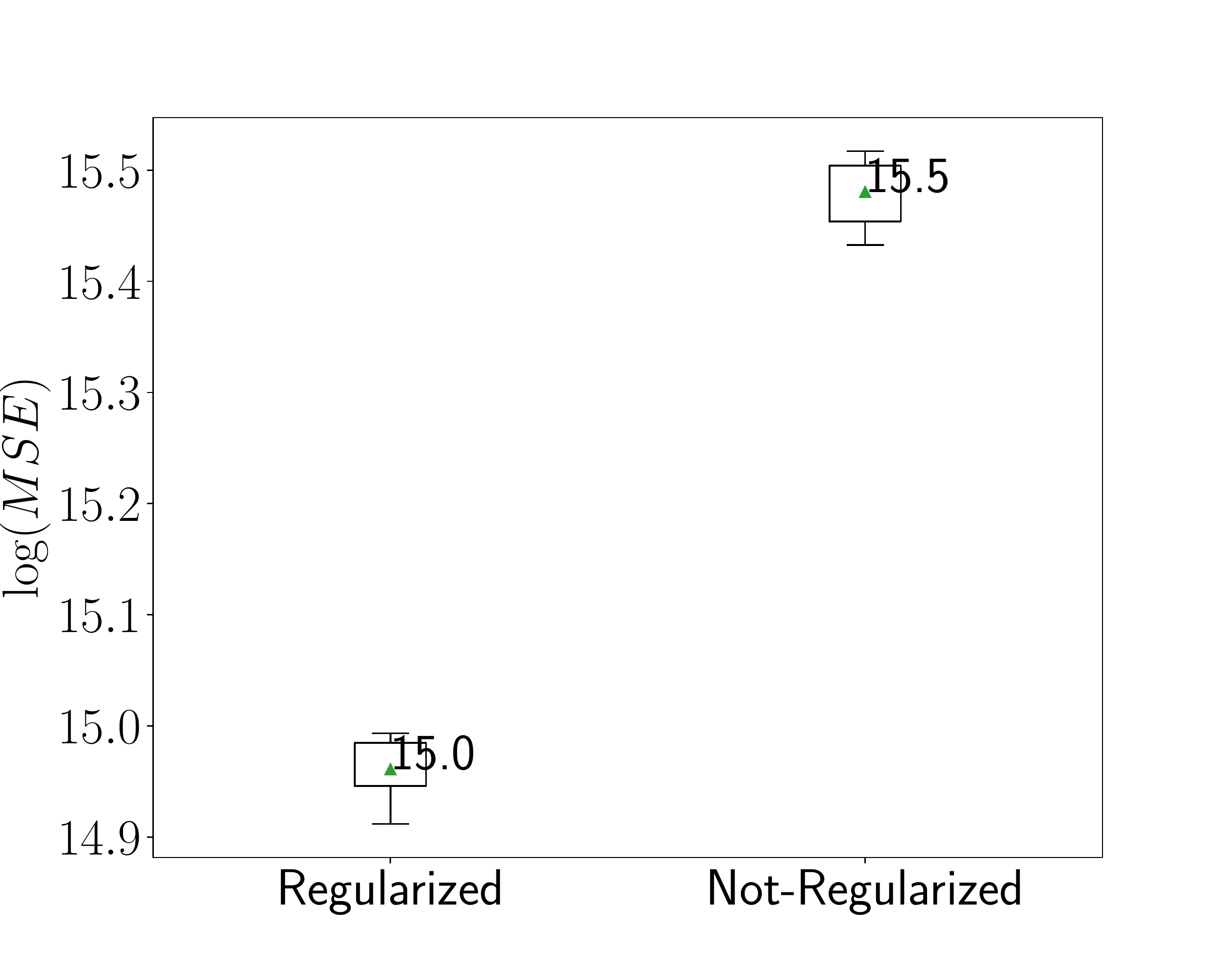} }
    \subfloat[\centering  ]{\includegraphics[width=0.33\textwidth]{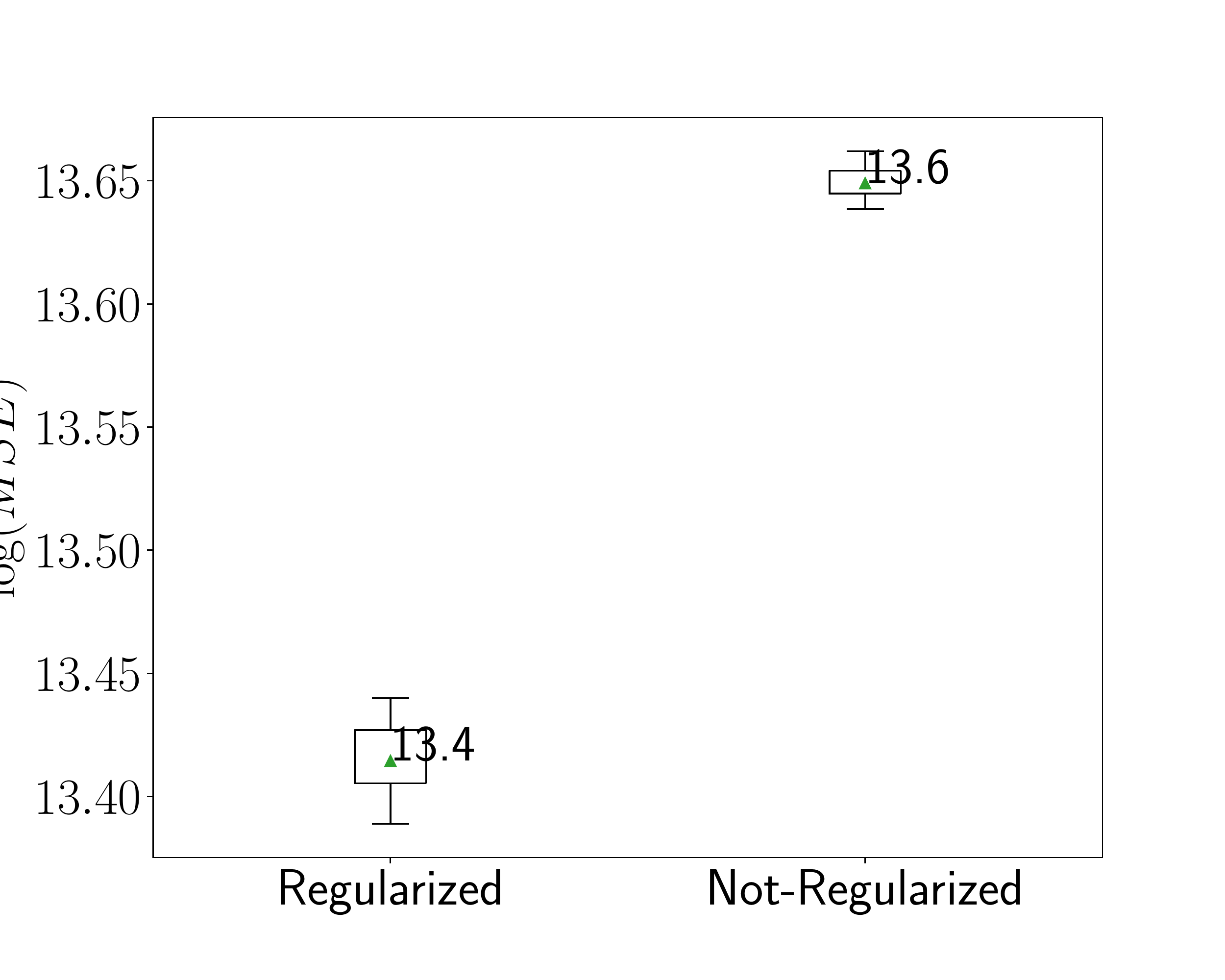} }
    \subfloat[\centering  ]{\includegraphics[width=0.33\textwidth]{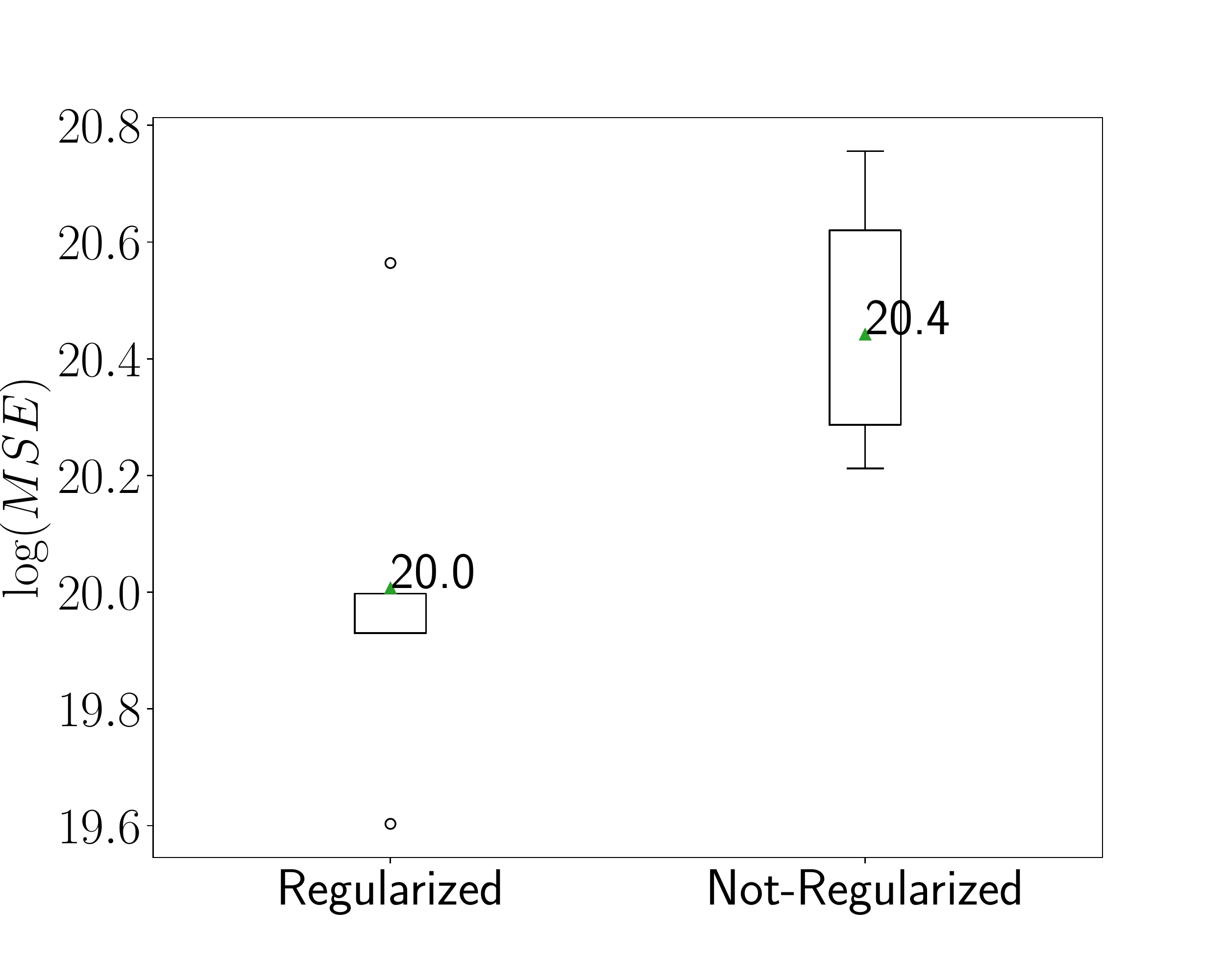} }
    \qquad
    \subfloat [\centering  ]{\includegraphics[width=0.33\textwidth]{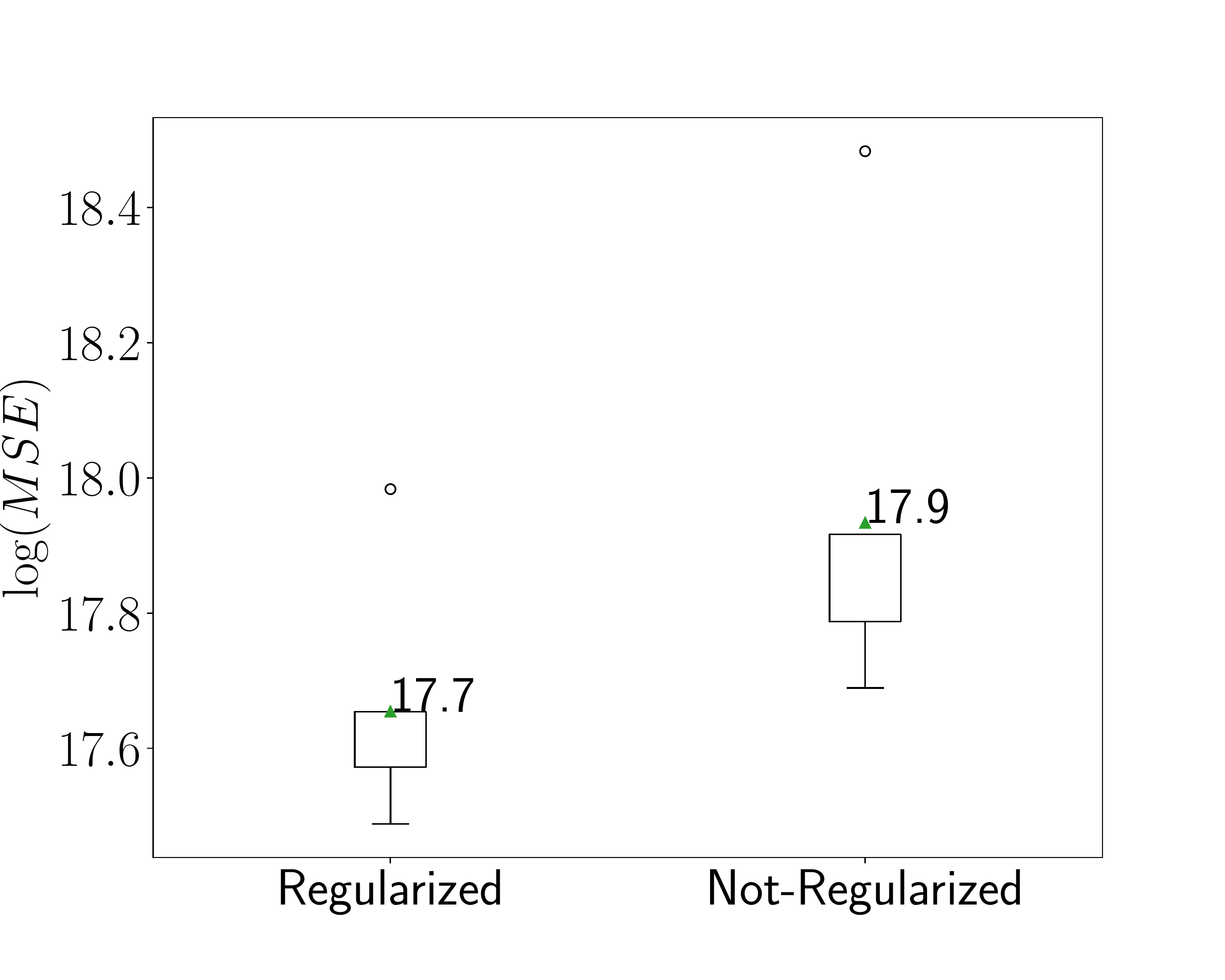} }
    \subfloat[\centering  ]{\includegraphics[width=0.33\textwidth]{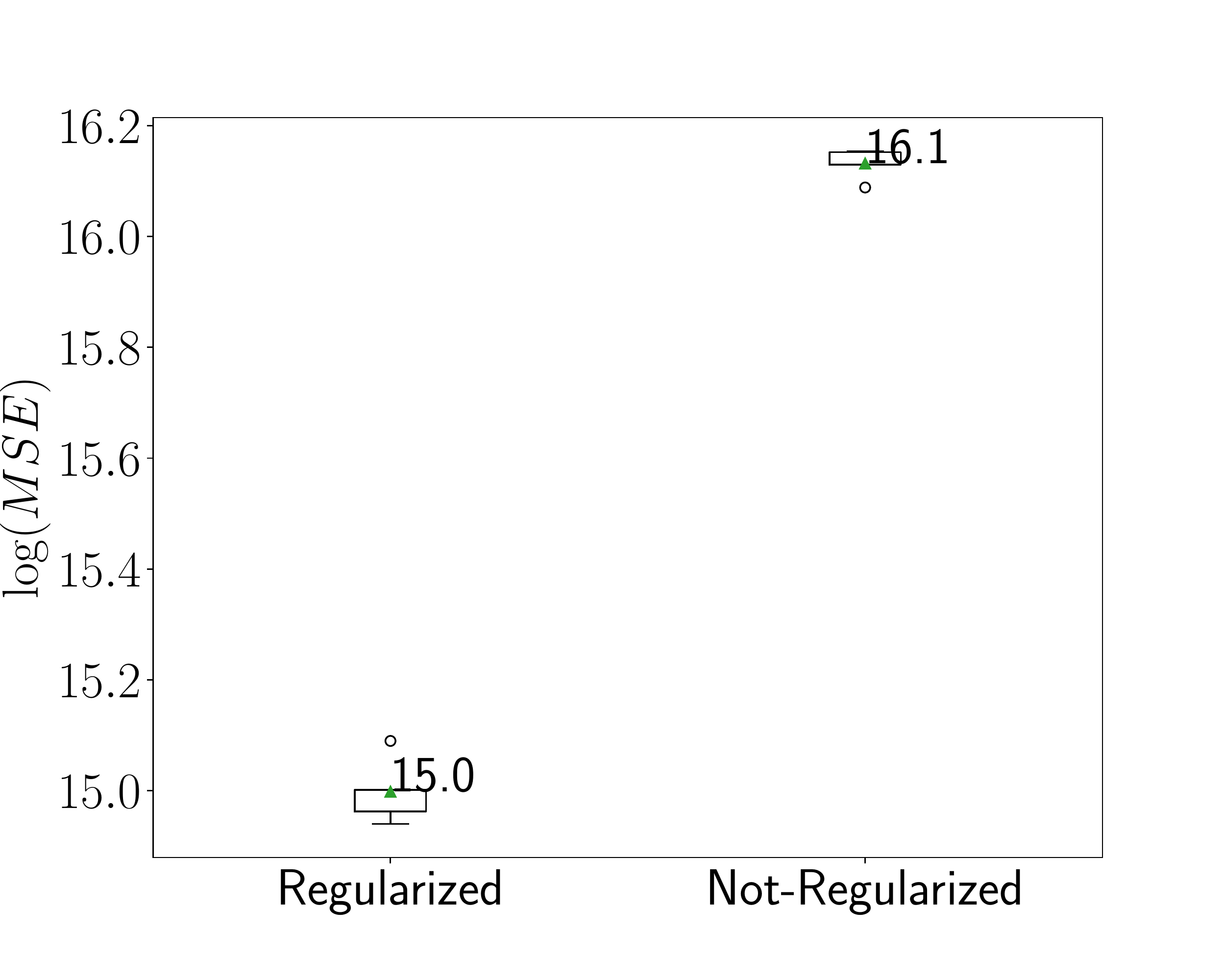} }
    \subfloat[\centering  ]{\includegraphics[width=0.33\textwidth]{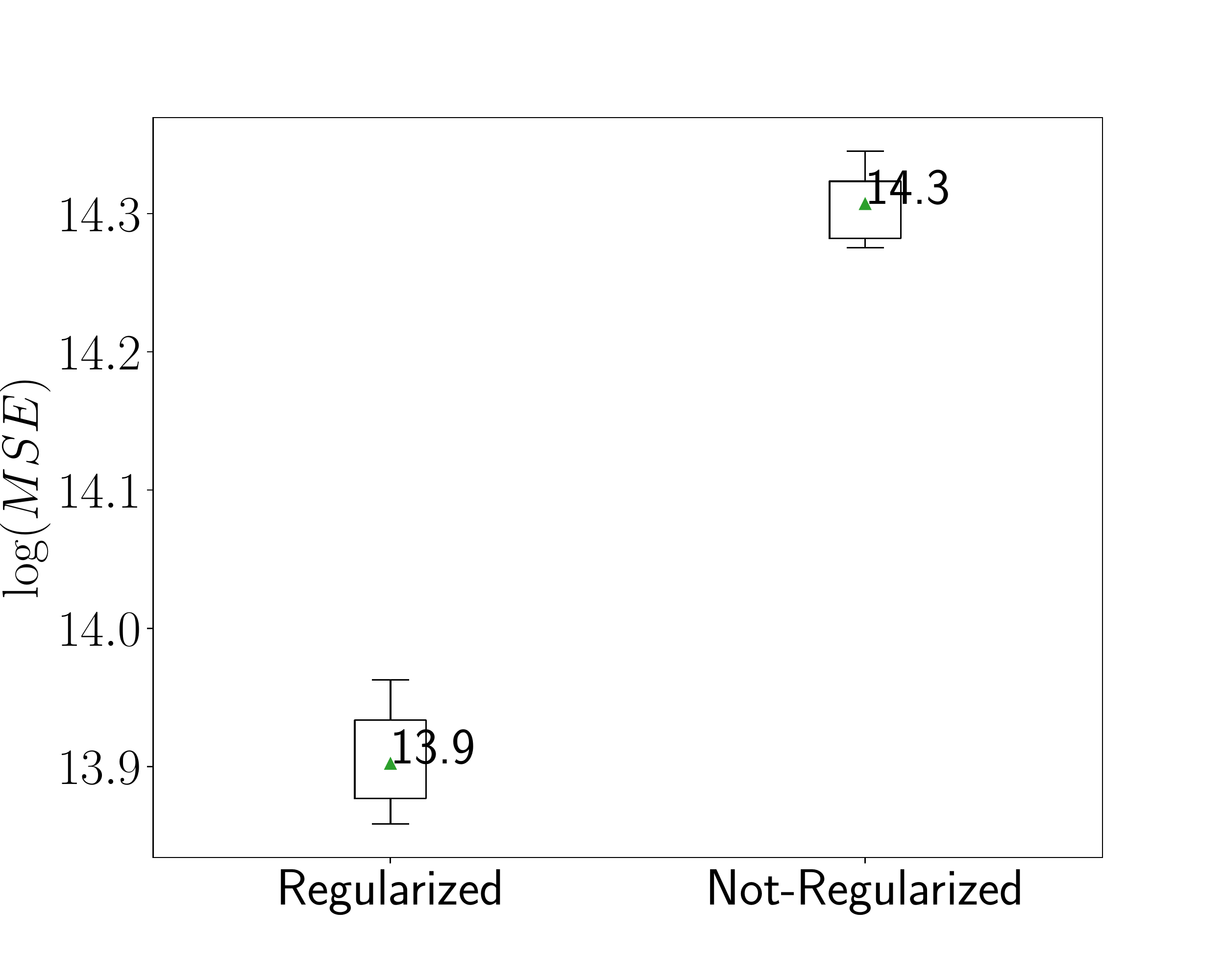} }
    \qquad
    \subfloat [\centering  ]{\includegraphics[width=0.33\textwidth]{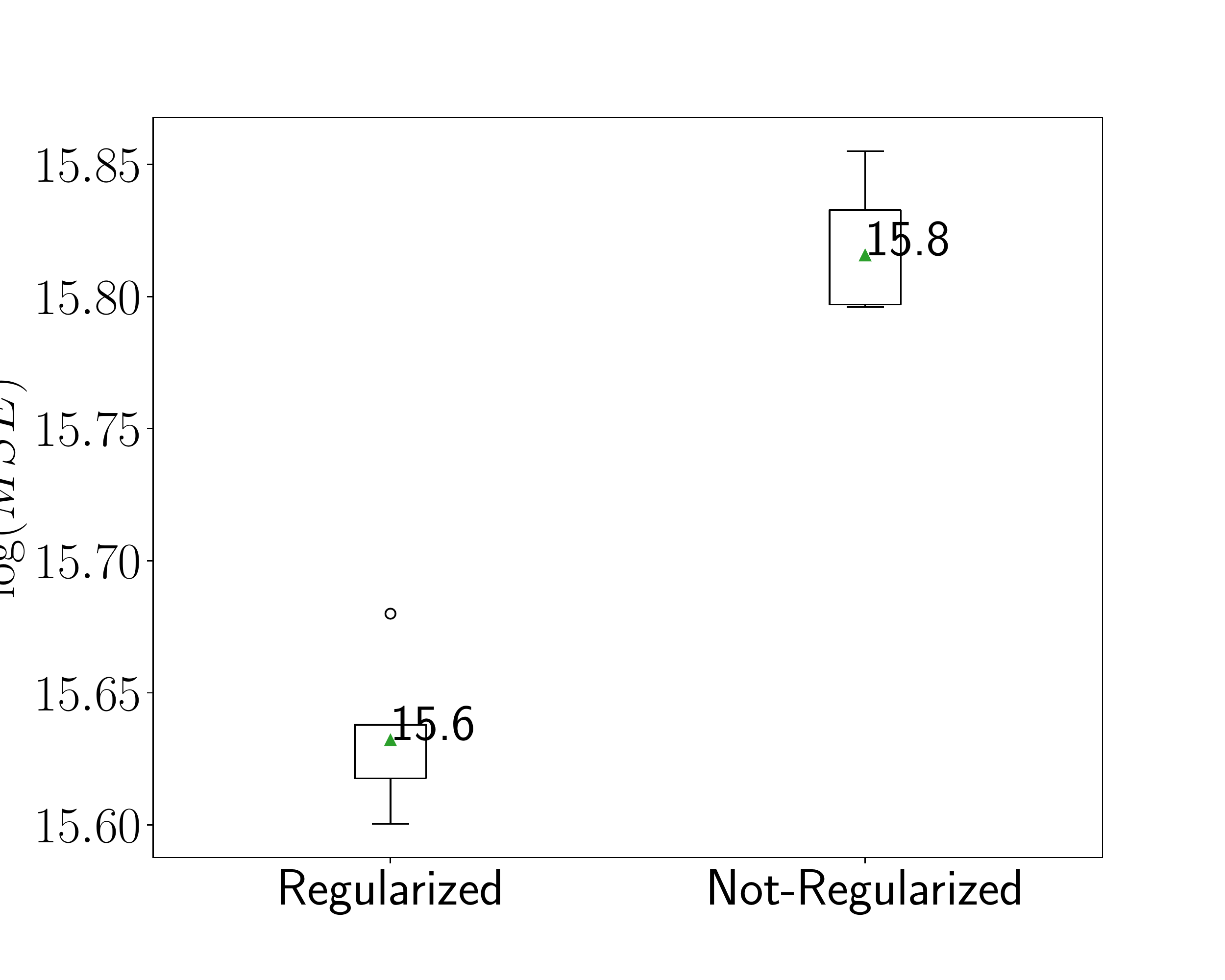} }
    \subfloat[\centering  ]{\includegraphics[width=0.33\textwidth]{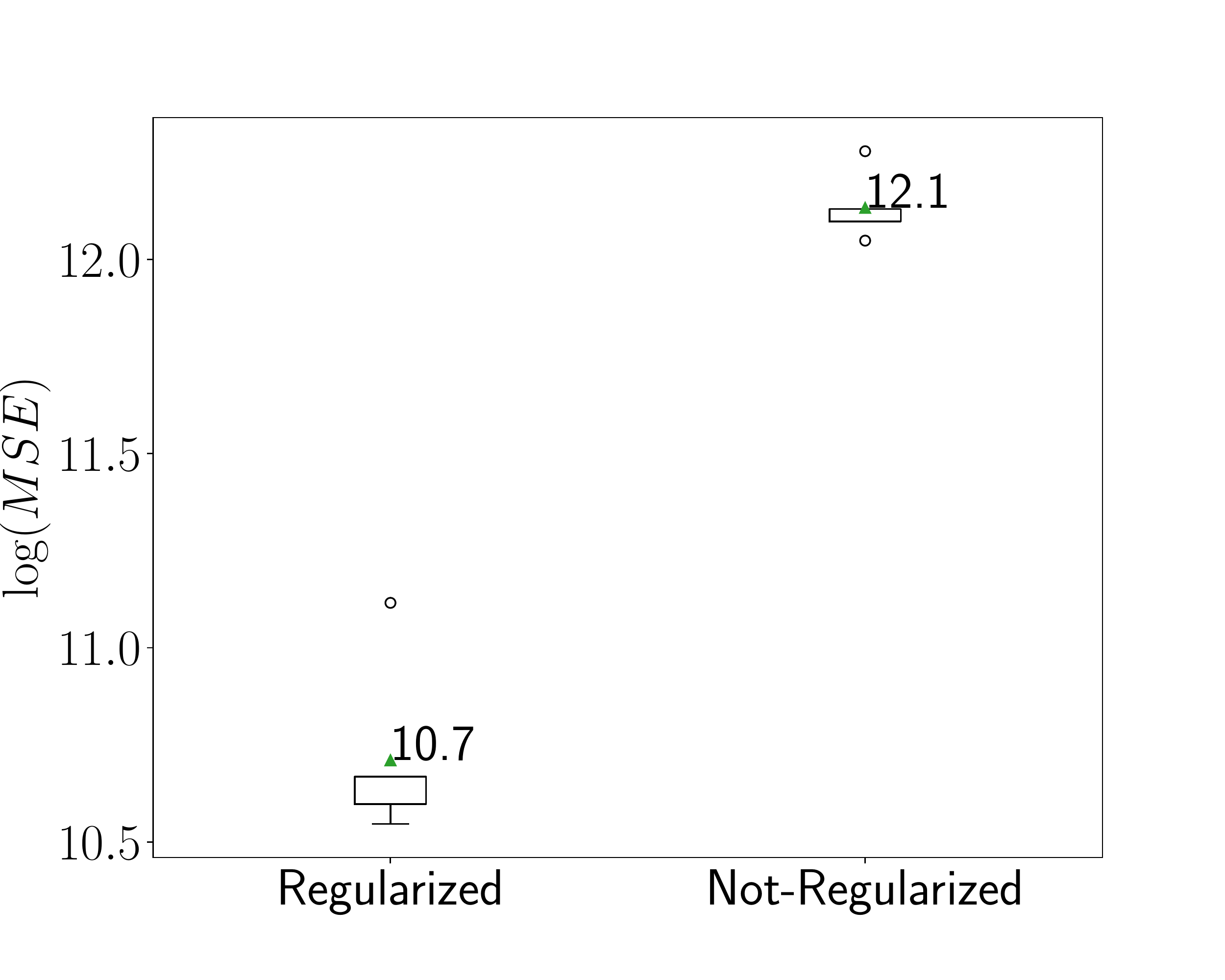} }
    \subfloat[\centering  ]{\includegraphics[width=0.33\textwidth]{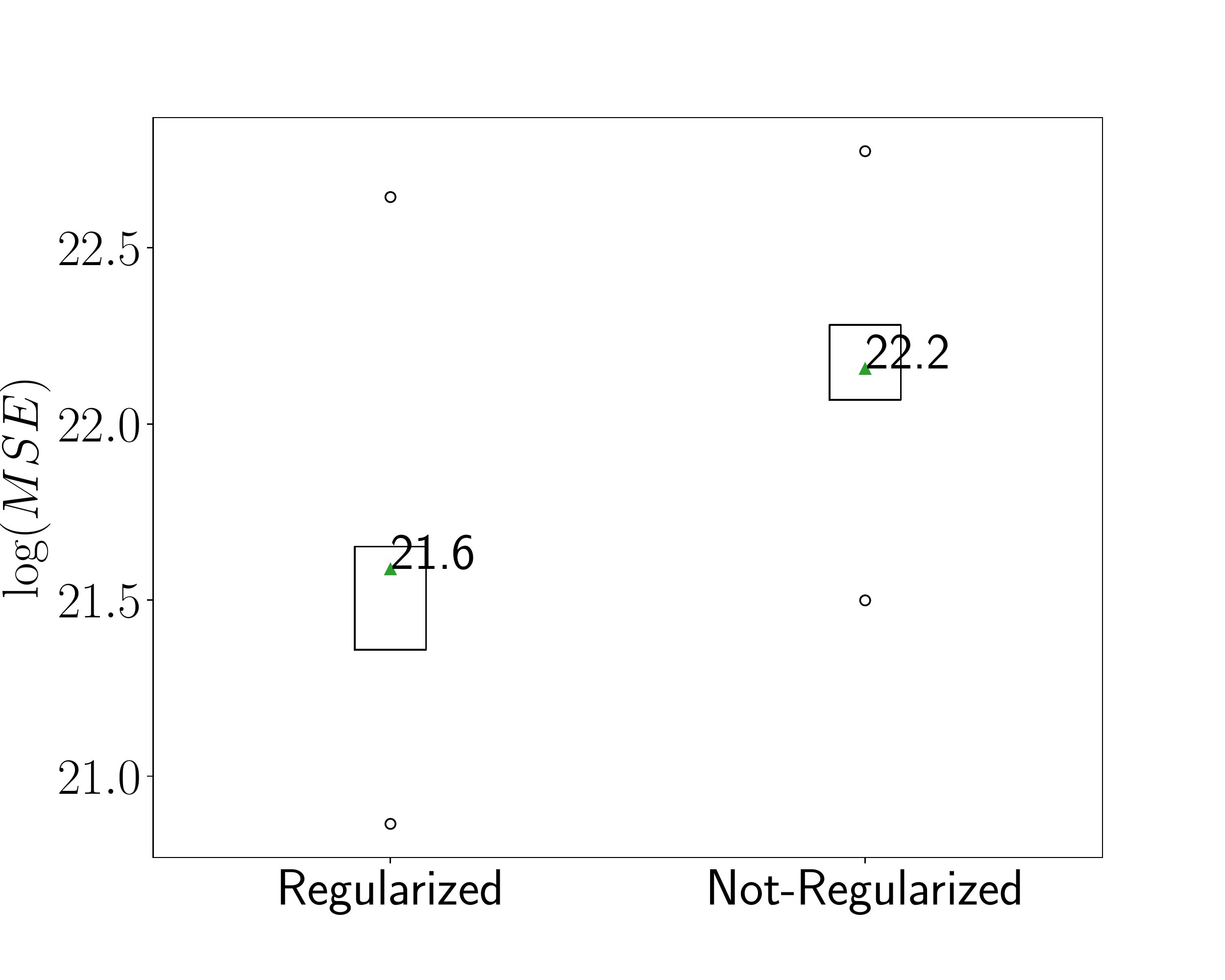} }
    \qquad
    \subfloat[\centering  ]{\includegraphics[width=0.33\textwidth]{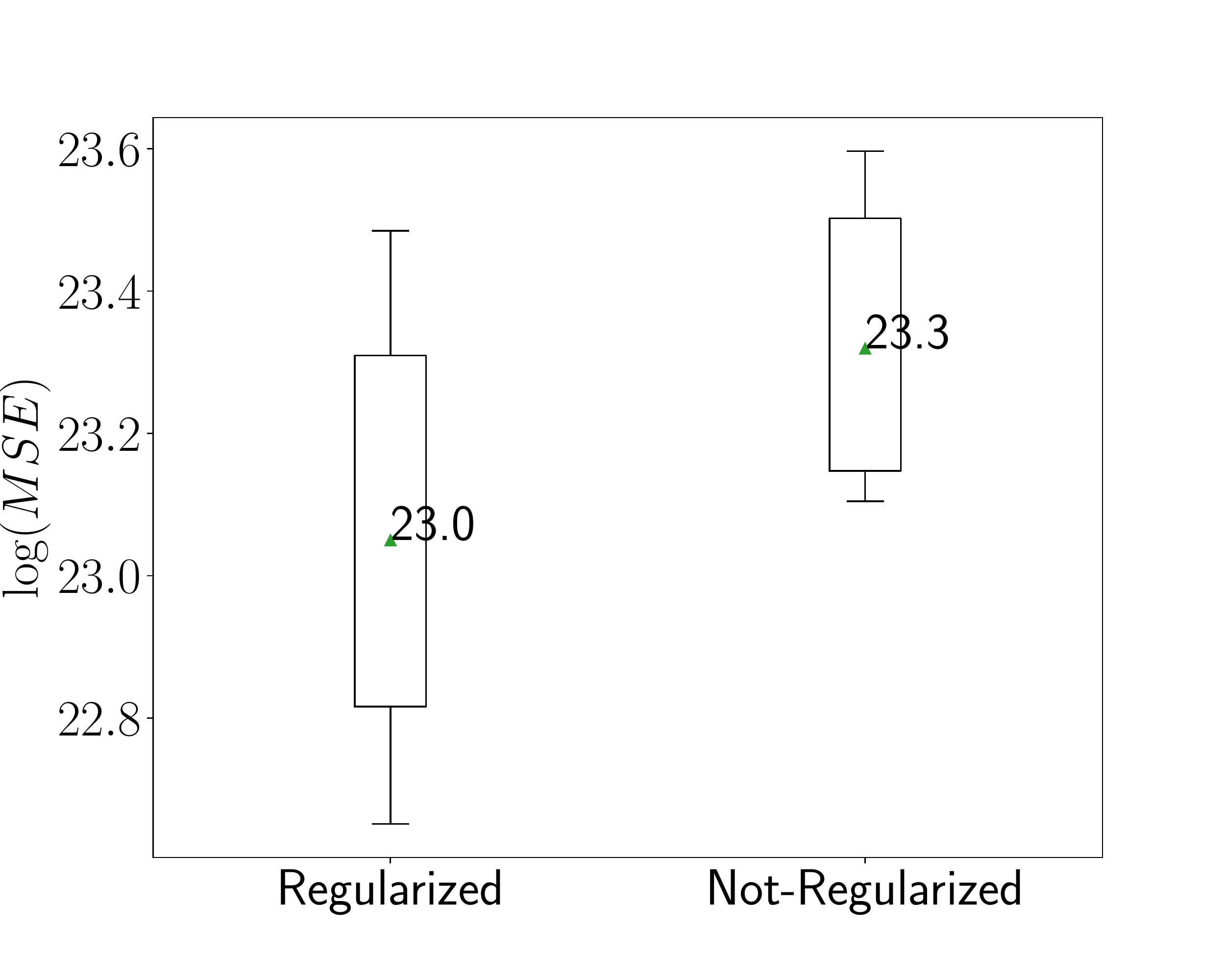} }
    \subfloat[\centering  ]{\includegraphics[width=0.33\textwidth]{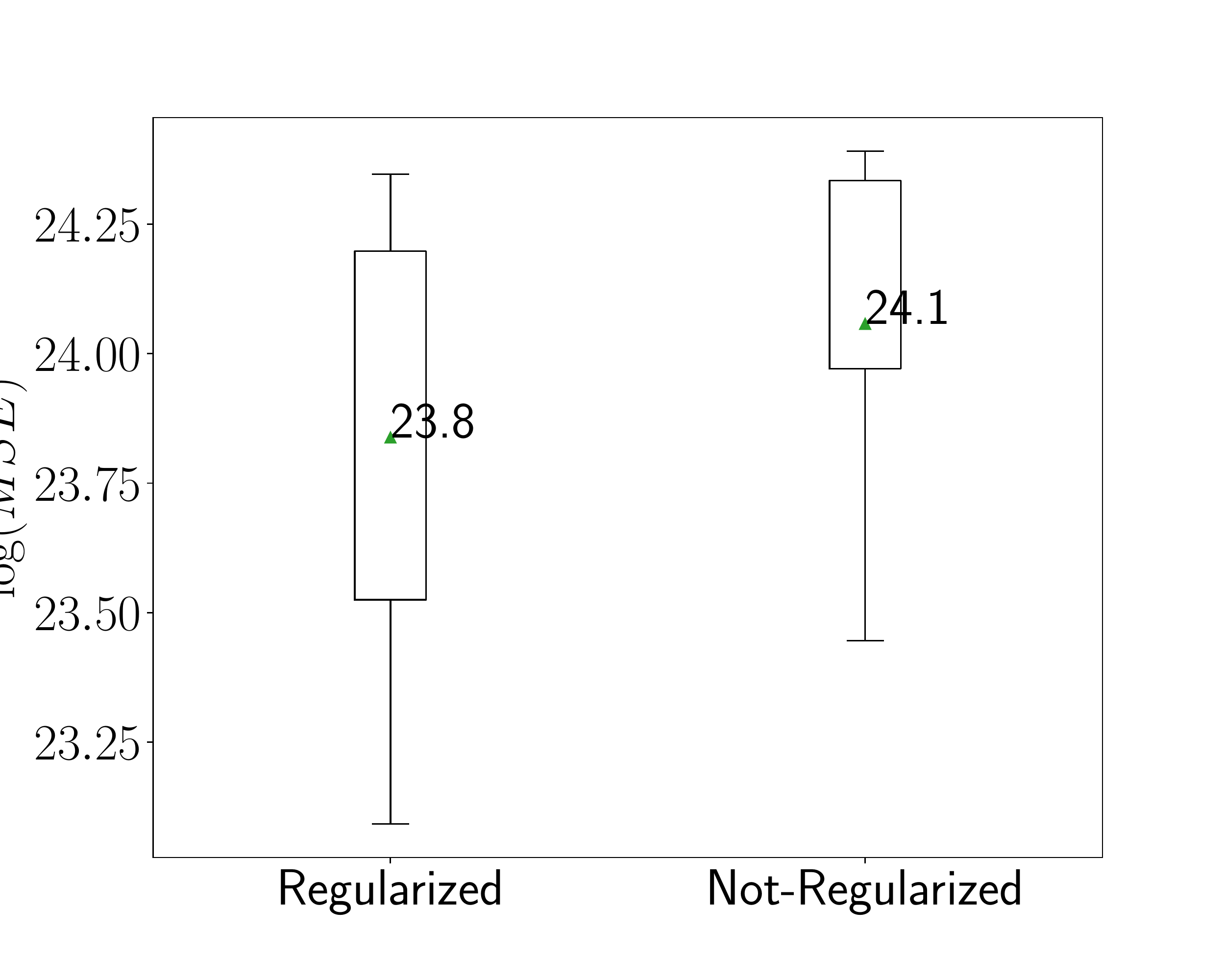} }
    \subfloat[\centering  ]{\includegraphics[width=0.33\textwidth]{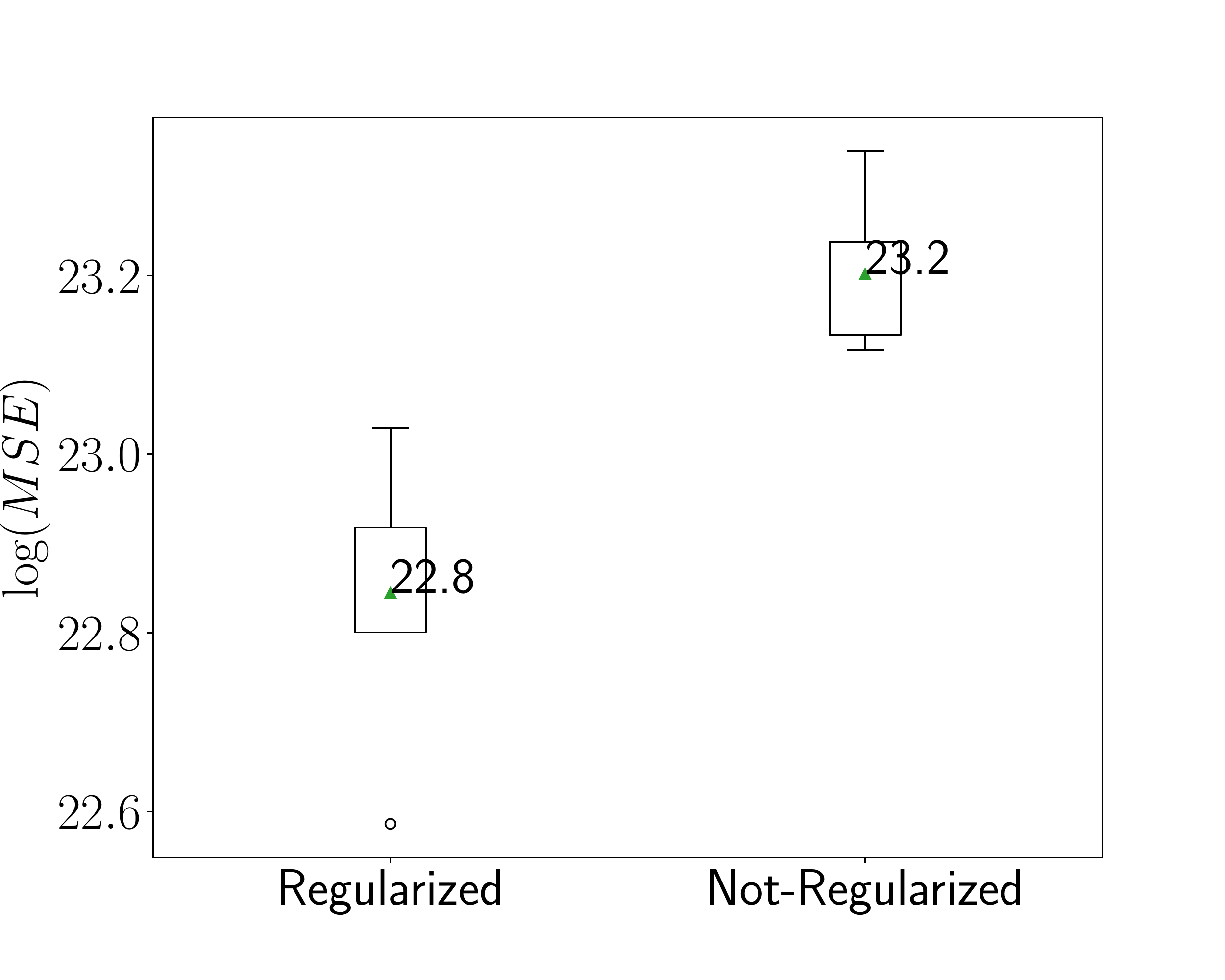} }
    \caption{The effect of regularization on DNMF performance in the supervised case. (a-l) represent simulated data sets (1-12), respectively.}
    \label{fig:compare reg}%
\end{figure}

We start with testing the different model formulations using simulated data.
First, we compare the regularized to the non-regularized variant in the supervised case.
As expected, the results, summarized in Figure~\ref{fig:compare reg}, show that the regularized
variant performs best in terms of MSE, hence we focus on it in the sequel.

\begin{figure}[H]
    \centering
    \subfloat [\centering Supervised ]{\includegraphics[width=0.45\textwidth]{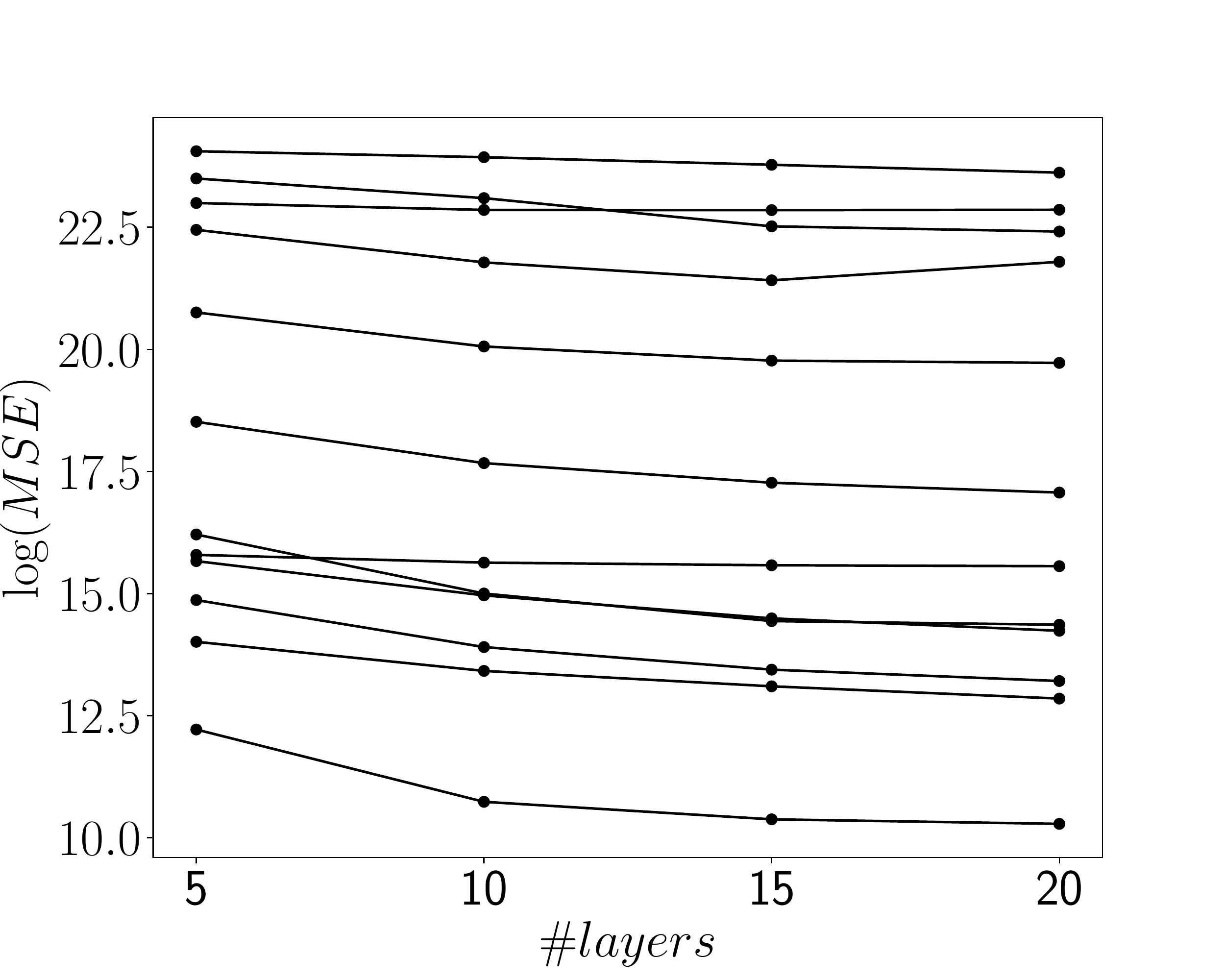} }
    \subfloat[\centering Unsupervised ]{\includegraphics[width=0.45\textwidth]{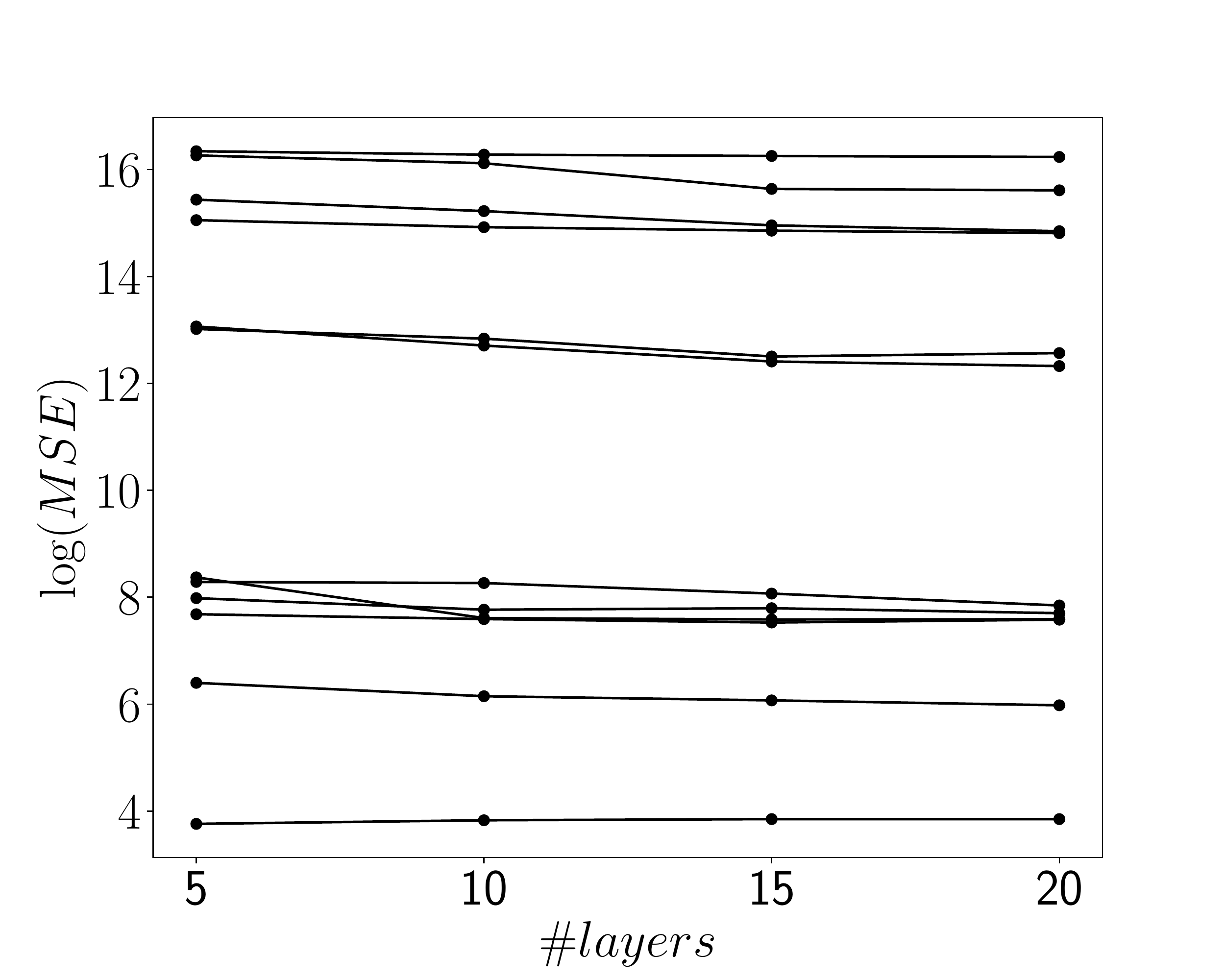} }
    \caption{The effect of number of layers on algorithm's performance. Each line corresponds to one of the simulated data sets.}
    \label{fig:depth}%
\end{figure}

Next, we tested the effect of the depth of the unrolled network on the algorithm's performance. 
The results, depicted in Figure~\ref{fig:depth}, show that after 10-15 layers the performance reaches a plateu, hence we focus in the following on depth-10 networks.
Notably, as our network borrows from the MU update scheme and does not rely on activation functions, it is less affected by the problem of gradient decay for deep architectures. 


    

\begin{figure}
    \centering
    \subfloat [\centering  ]{\includegraphics[width=0.33\textwidth]{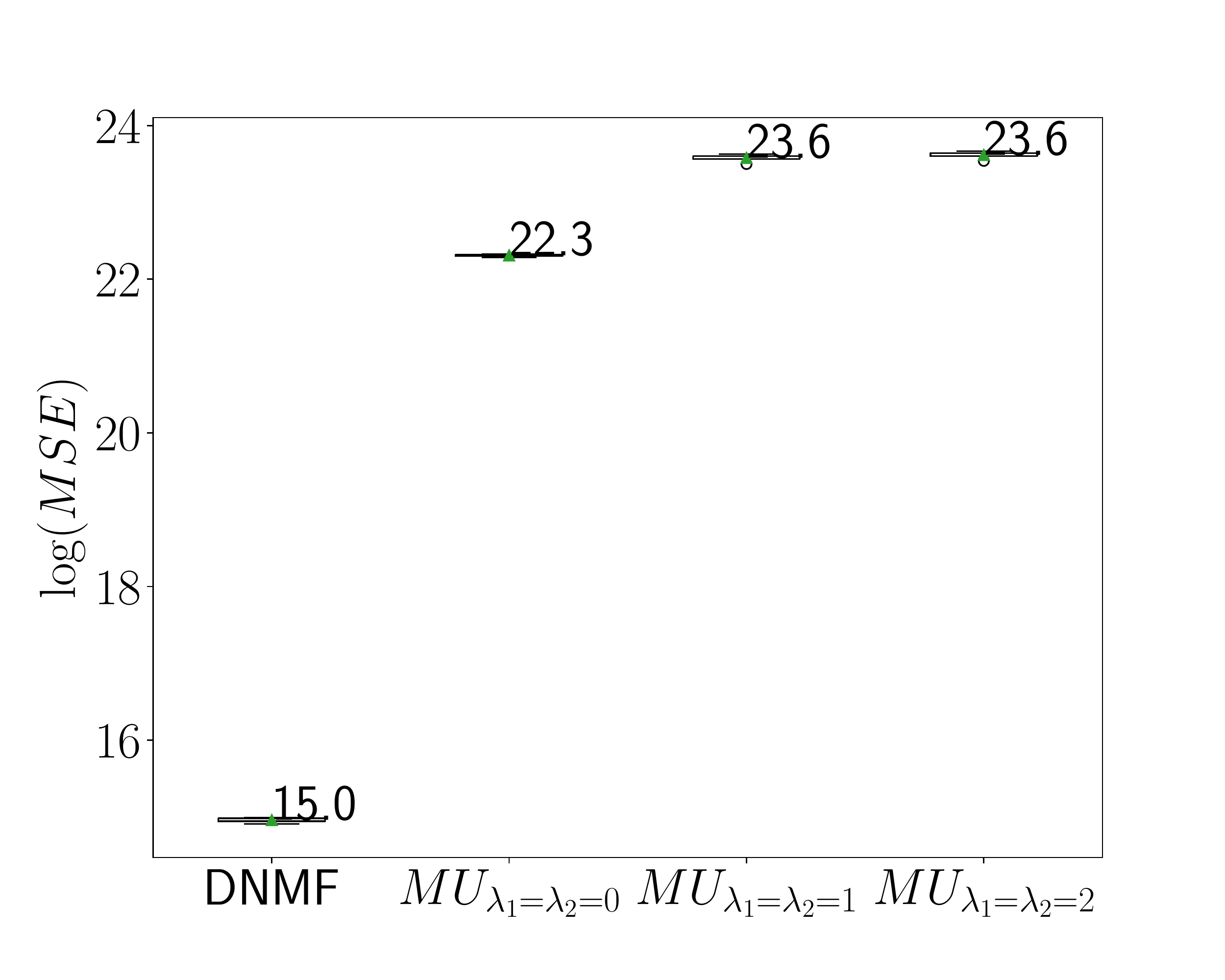} }
    \subfloat[\centering  ]{\includegraphics[width=0.33\textwidth]{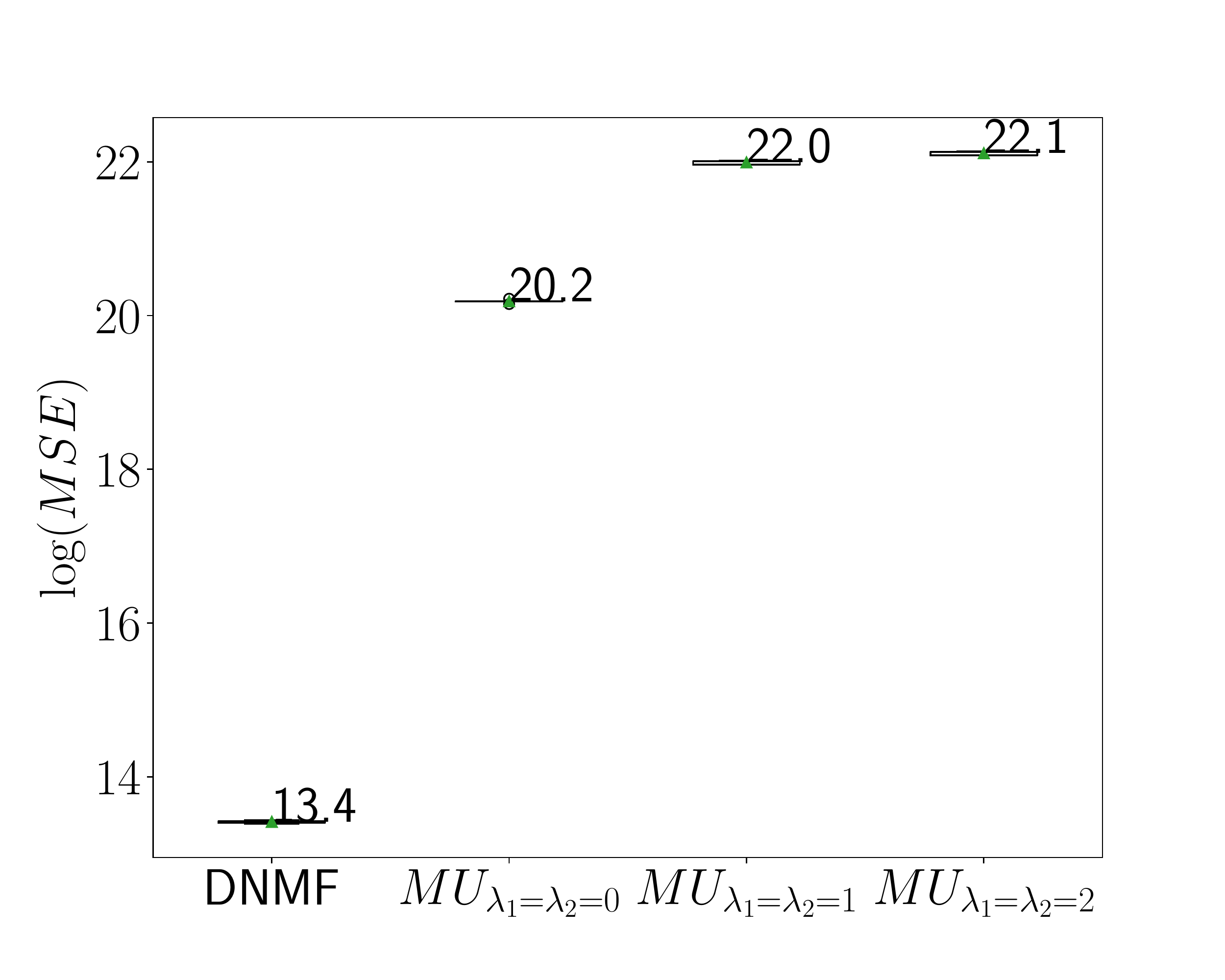} }
    \subfloat[\centering  ]{\includegraphics[width=0.33\textwidth]{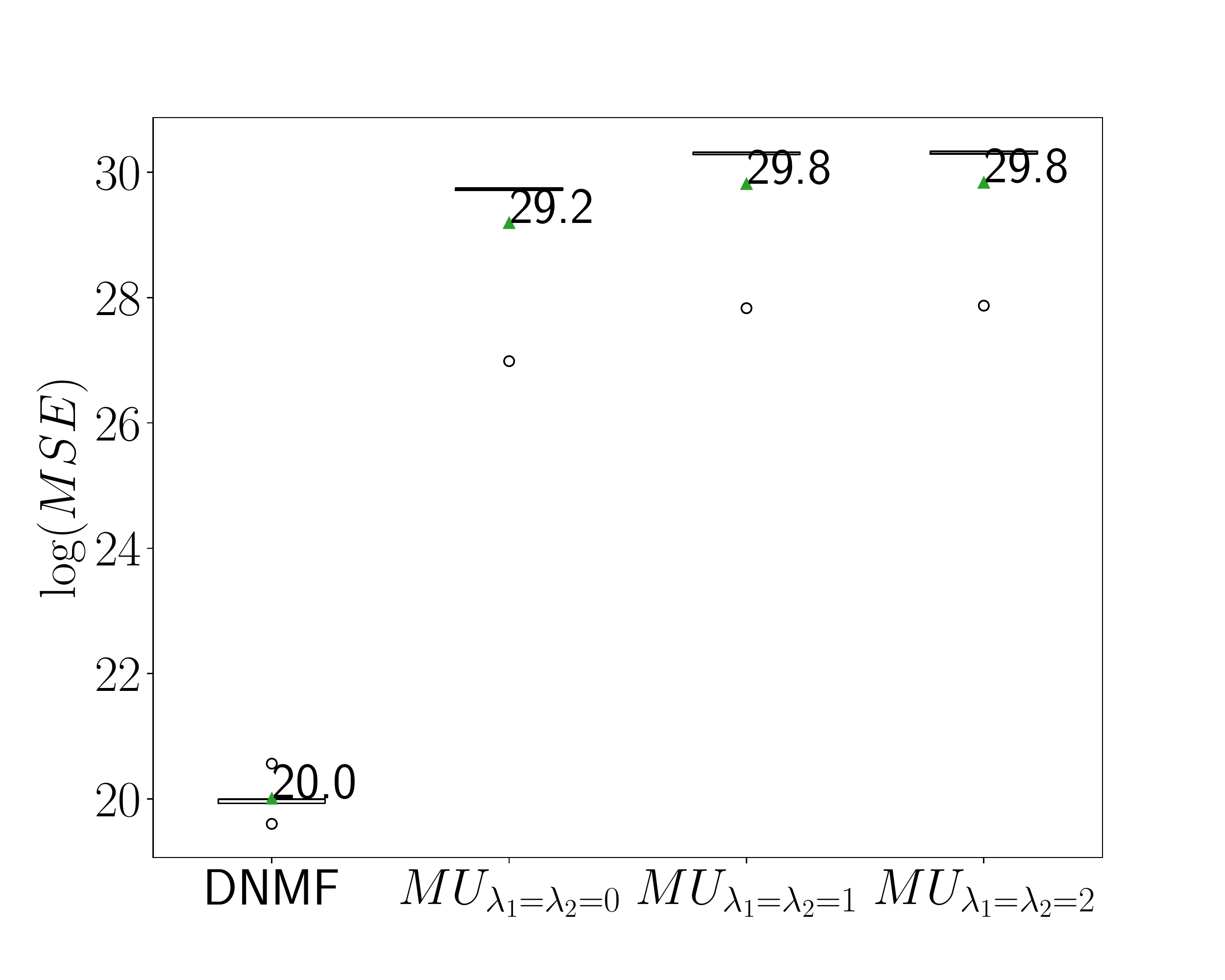} }
    \qquad
    \subfloat [\centering  ]{\includegraphics[width=0.33\textwidth]{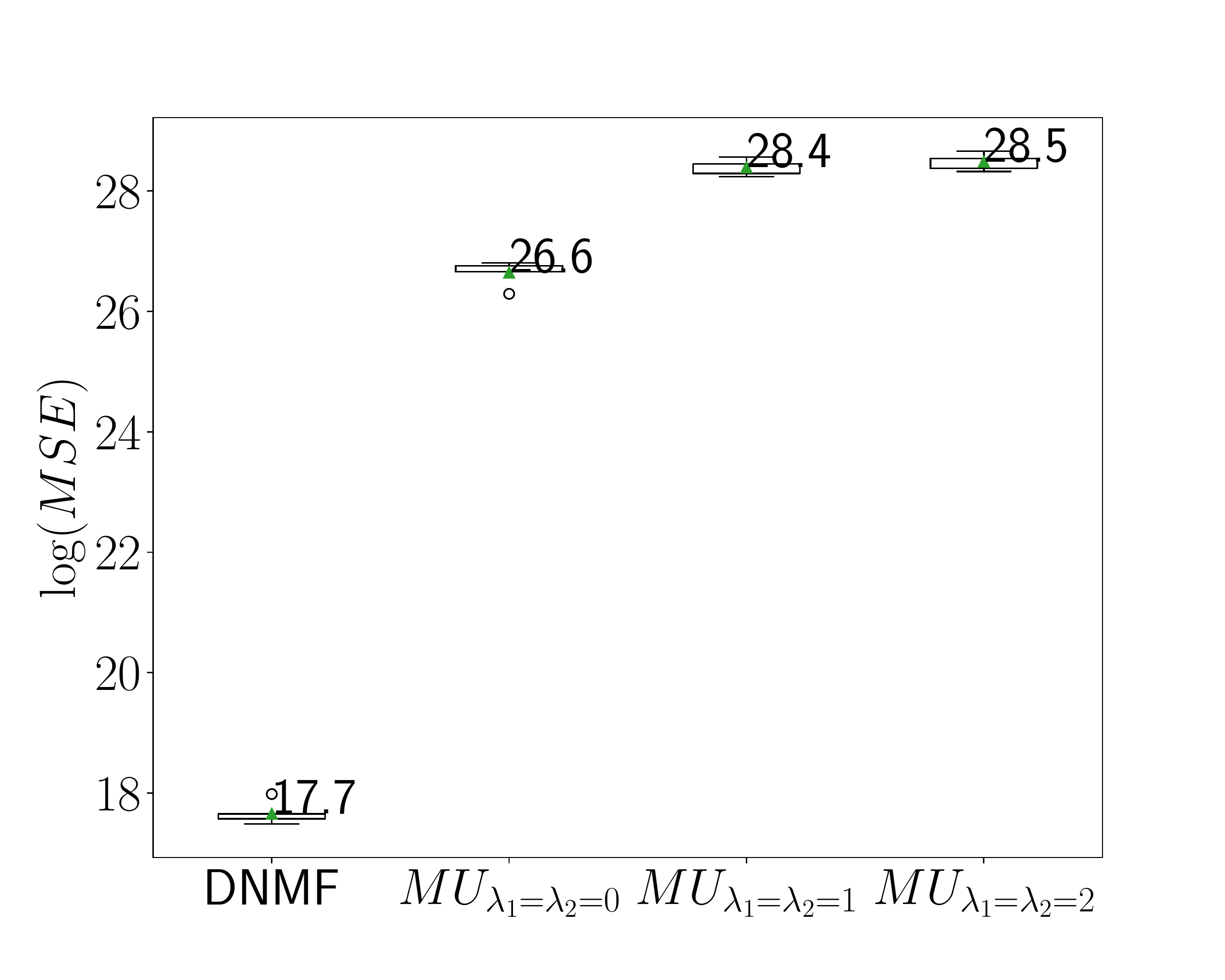} }
    \subfloat[\centering  ]{\includegraphics[width=0.33\textwidth]{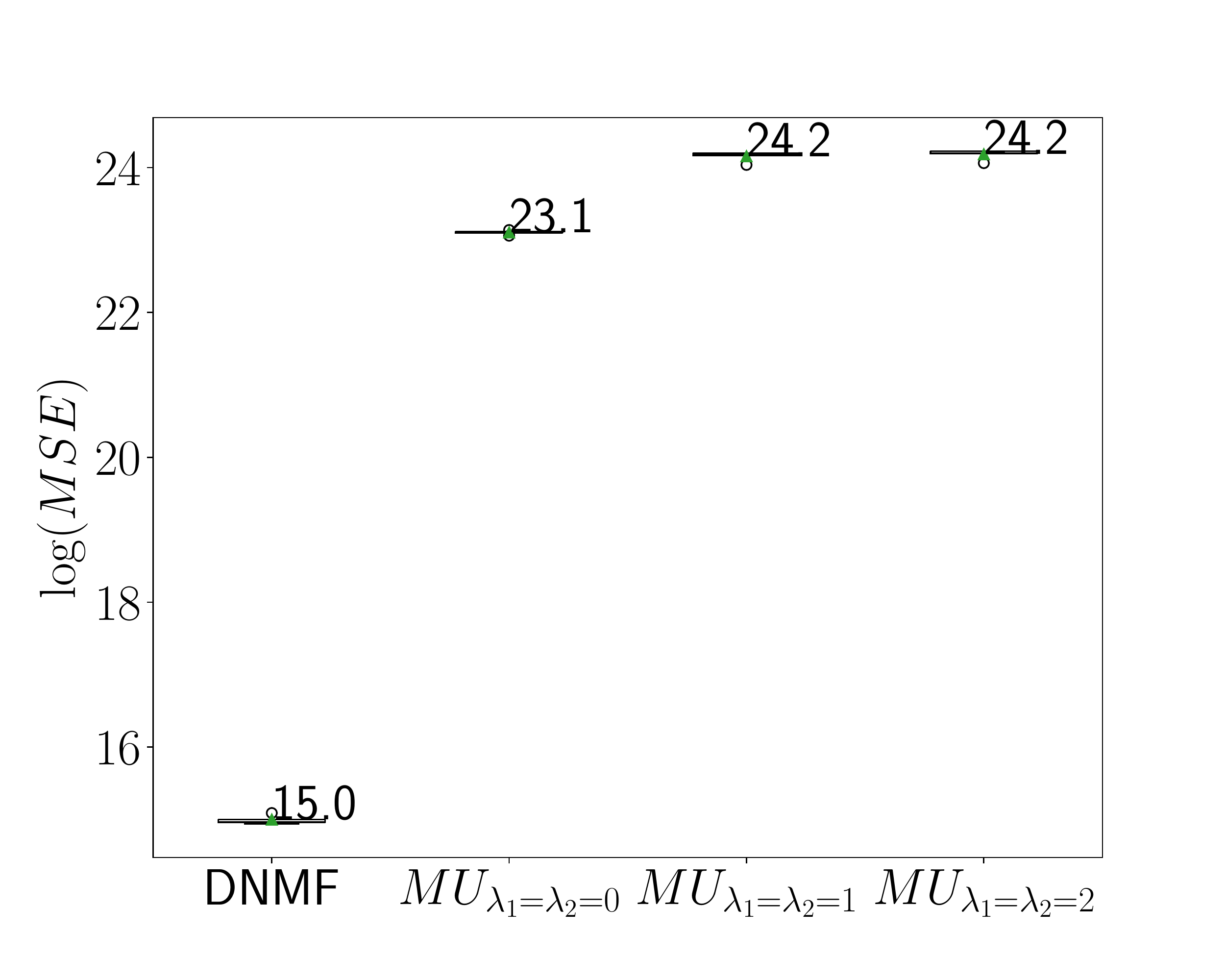} }
    \subfloat[\centering  ]{\includegraphics[width=0.33\textwidth]{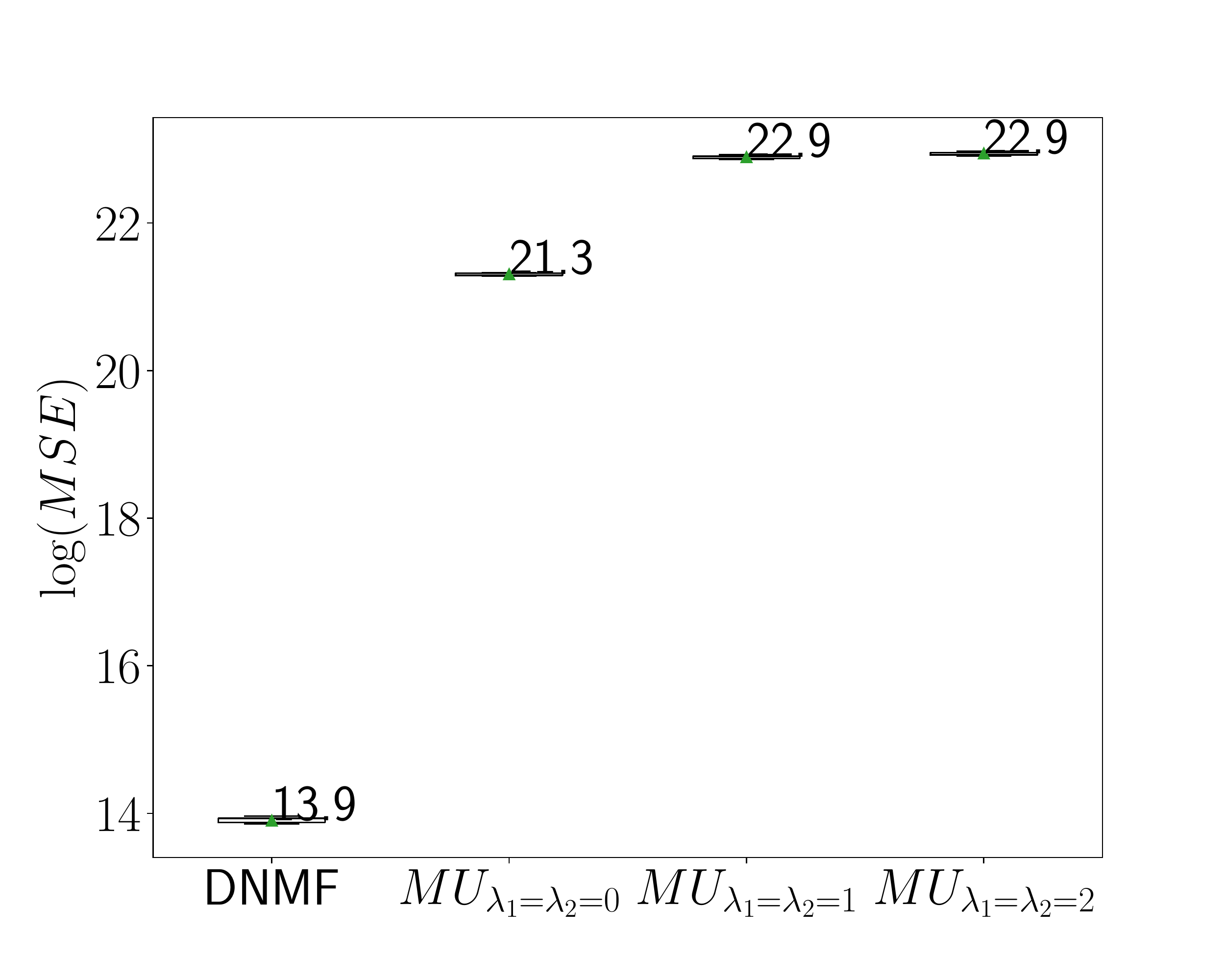} }
    \qquad
    \subfloat [\centering  ]{\includegraphics[width=0.33\textwidth]{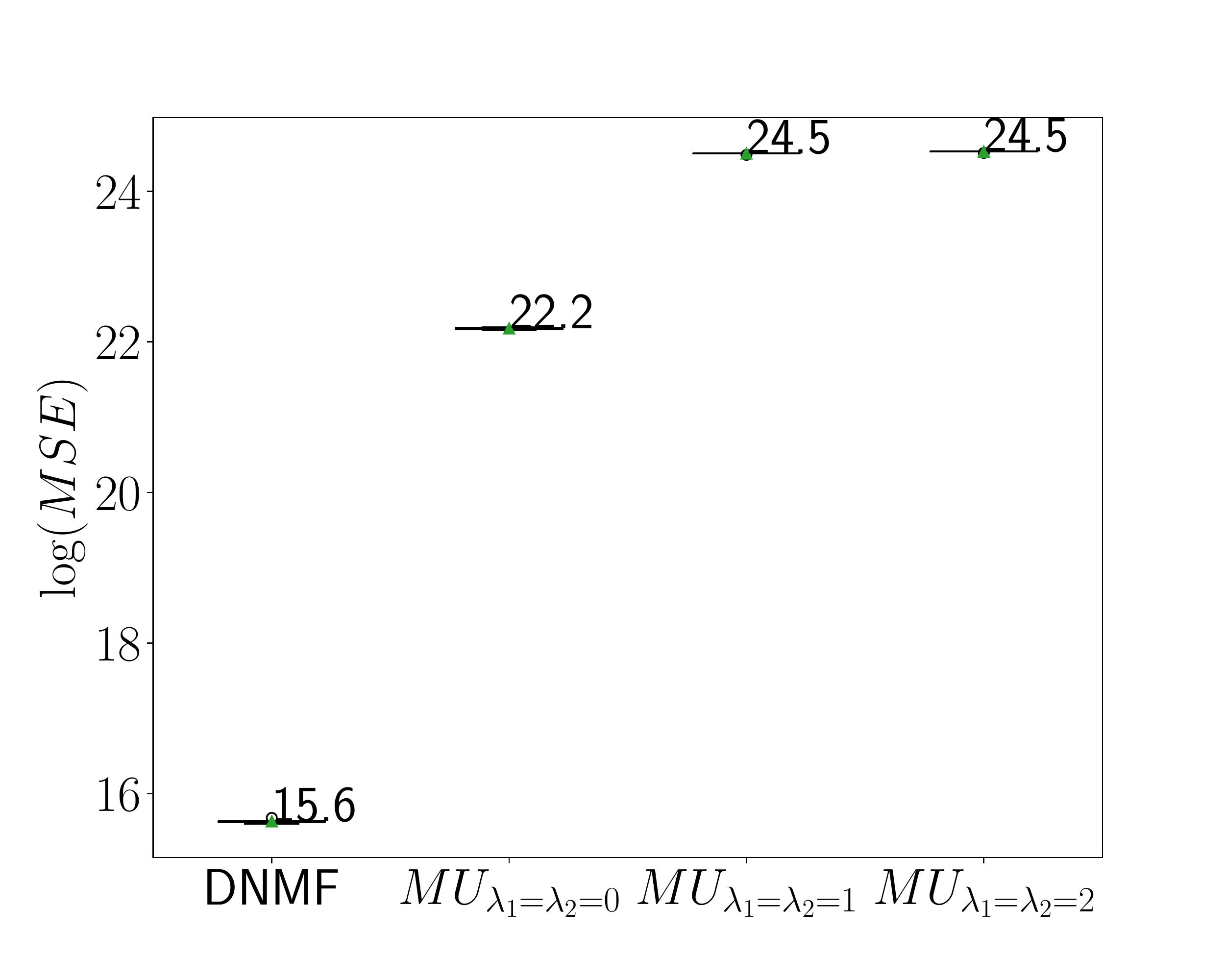} }
    \subfloat[\centering  ]{\includegraphics[width=0.33\textwidth]{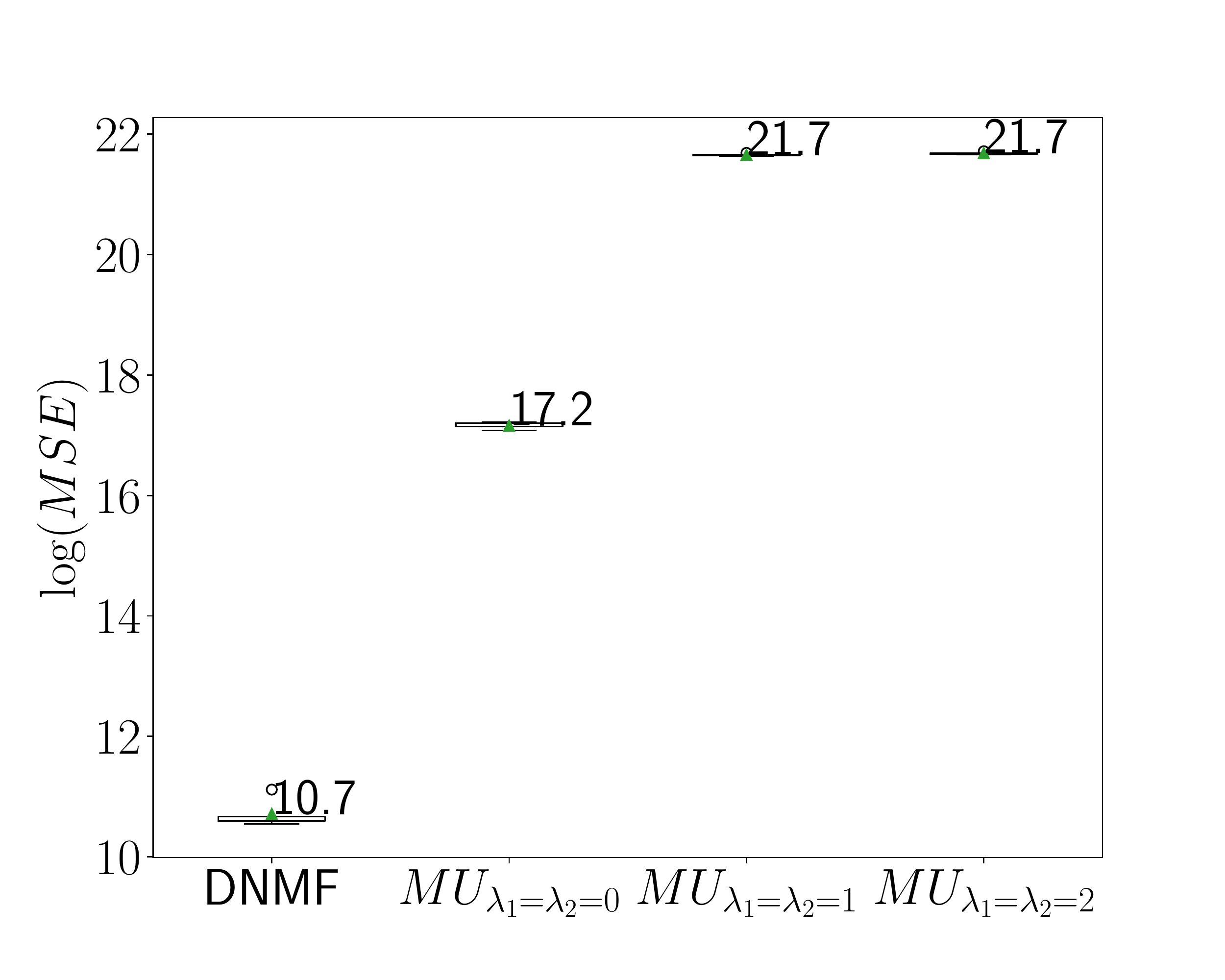} }
    \subfloat[\centering  ]{\includegraphics[width=0.33\textwidth]{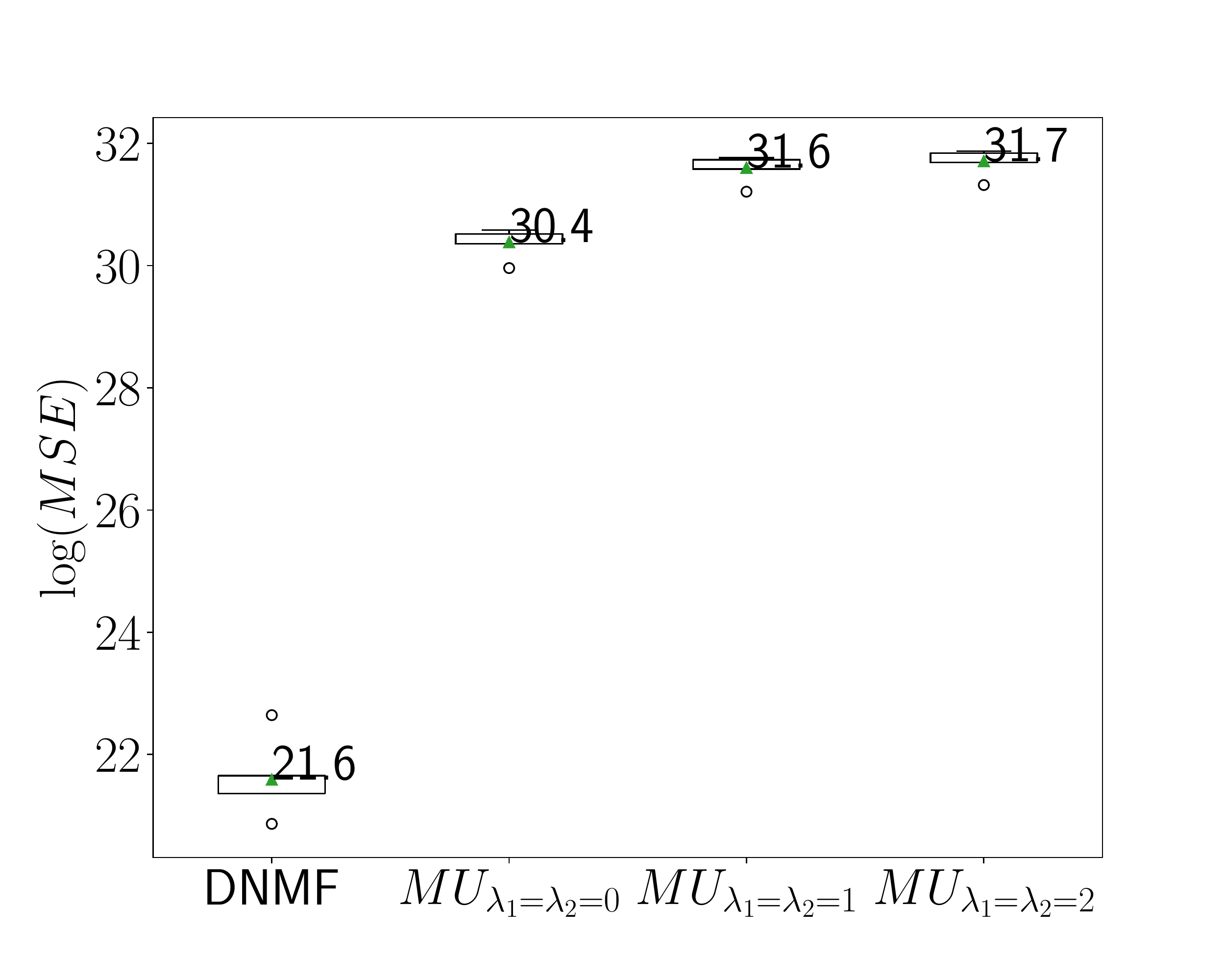} }
    \qquad
    \subfloat[\centering  ]{\includegraphics[width=0.33\textwidth]{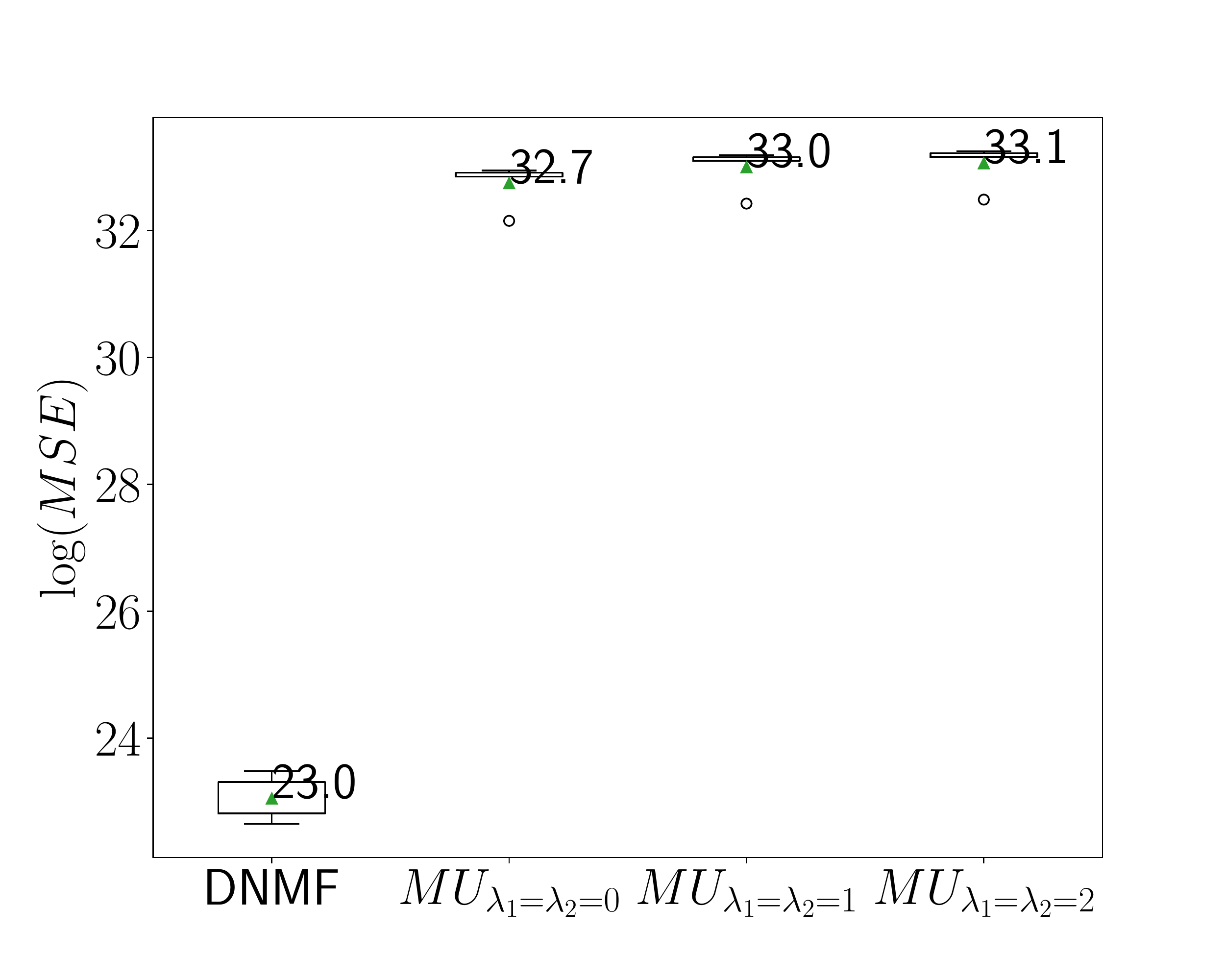} }
    \subfloat[\centering  ]{\includegraphics[width=0.33\textwidth]{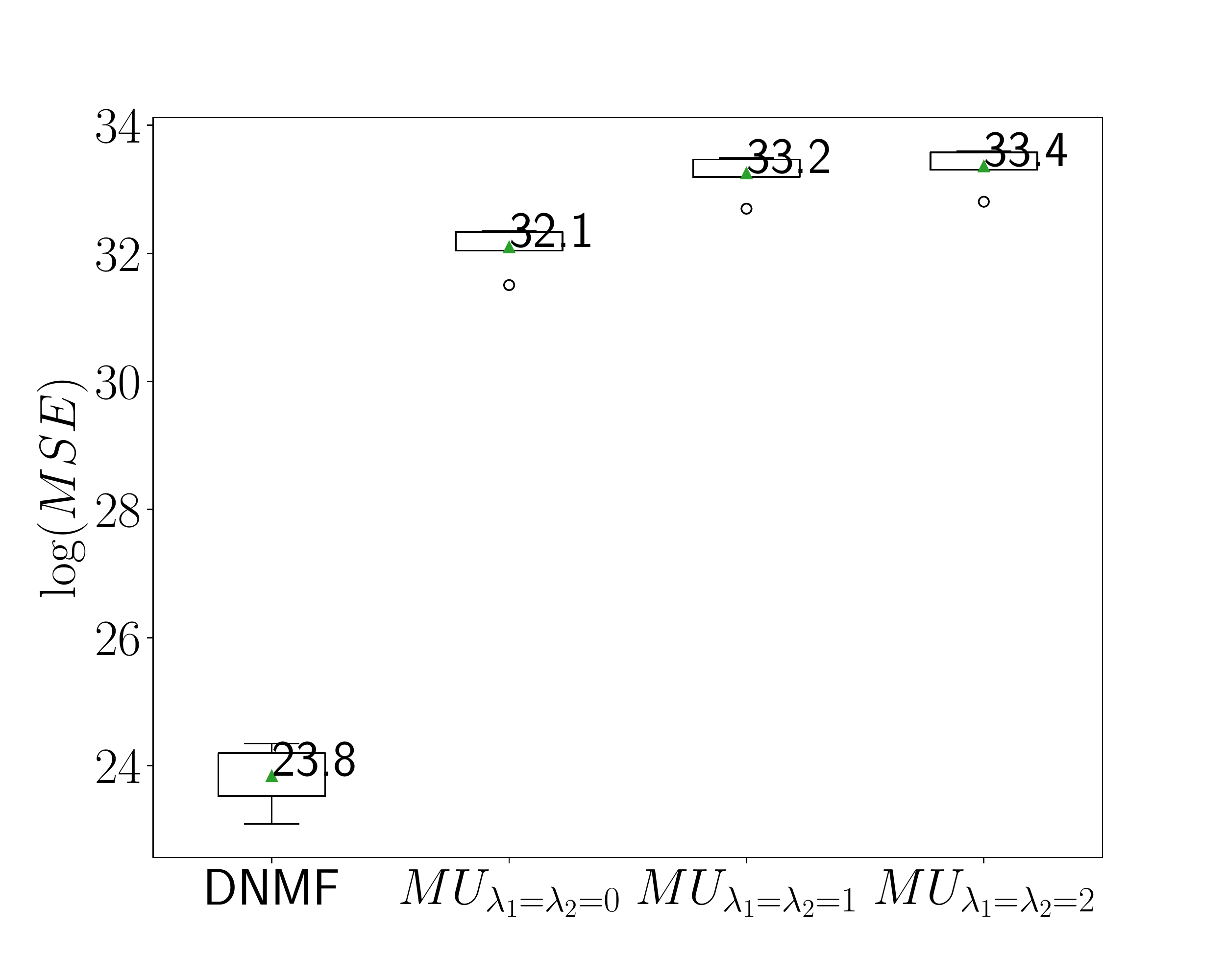} }
    \subfloat[\centering  ]{\includegraphics[width=0.33\textwidth]{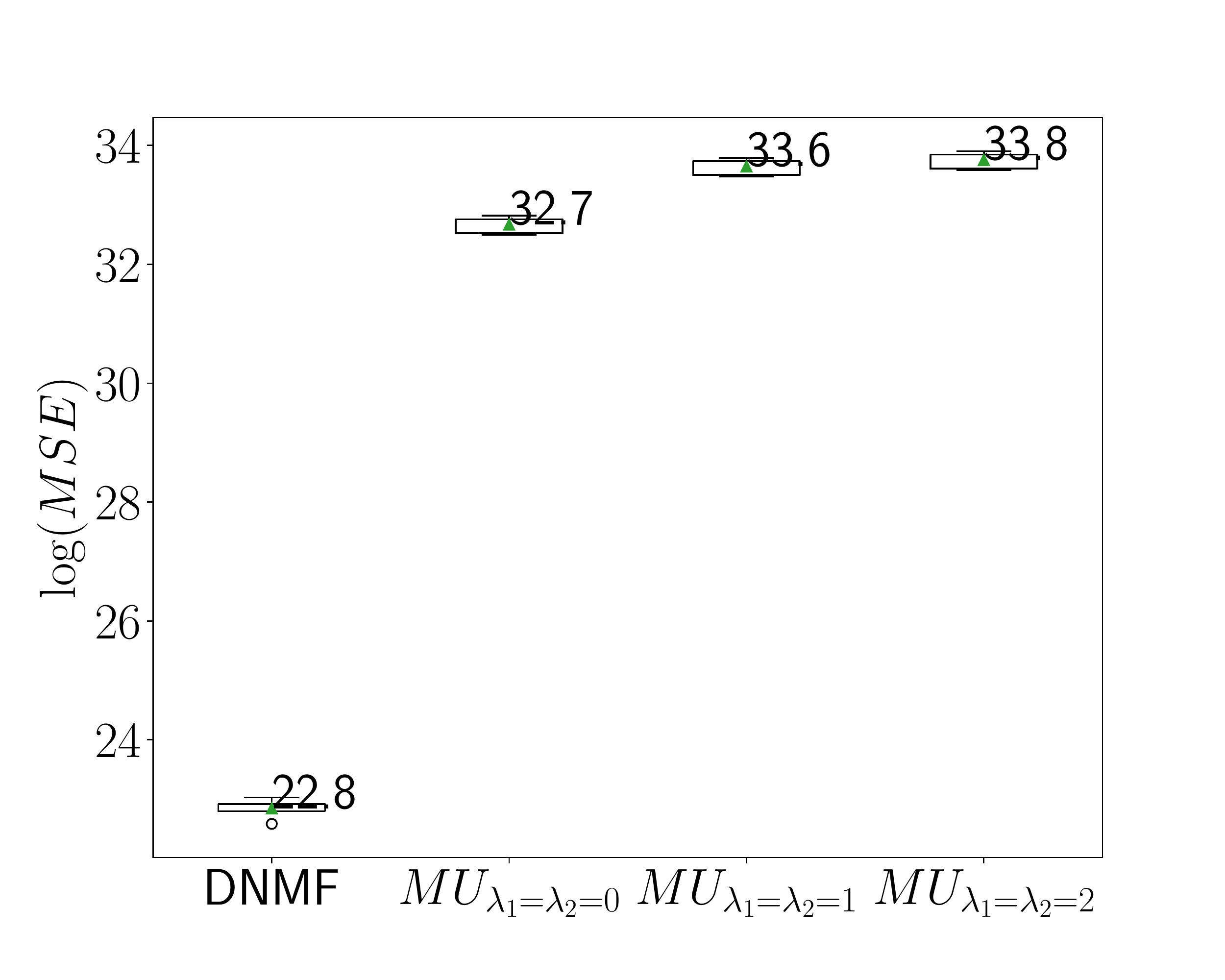} }
    \caption{Comparative performance on simulated data in the supervised setting. (a-l) represent simulated data sets (1-12), respectively.}
    \label{fig: Supervised DNMF-MU}%
\end{figure}

After determining the architecture of the developed framework, we turn to examine it in the supervised
case and compare to the MU approach on the simulated data. To this end, we apply MU to the training data to estimate $W$, and
then use MU with the learned $W$ to estimate $H$ on the test data. The results, summarized in Figure~\ref{fig: Supervised DNMF-MU}, show that DNMF outperforms MU across a wide range of regularization values for the latter (note that DNMF learns the regularization parameters automatically from data in this case).


\begin{figure}
    \centering
    \subfloat [\centering  ]{\includegraphics[width=0.33\textwidth]{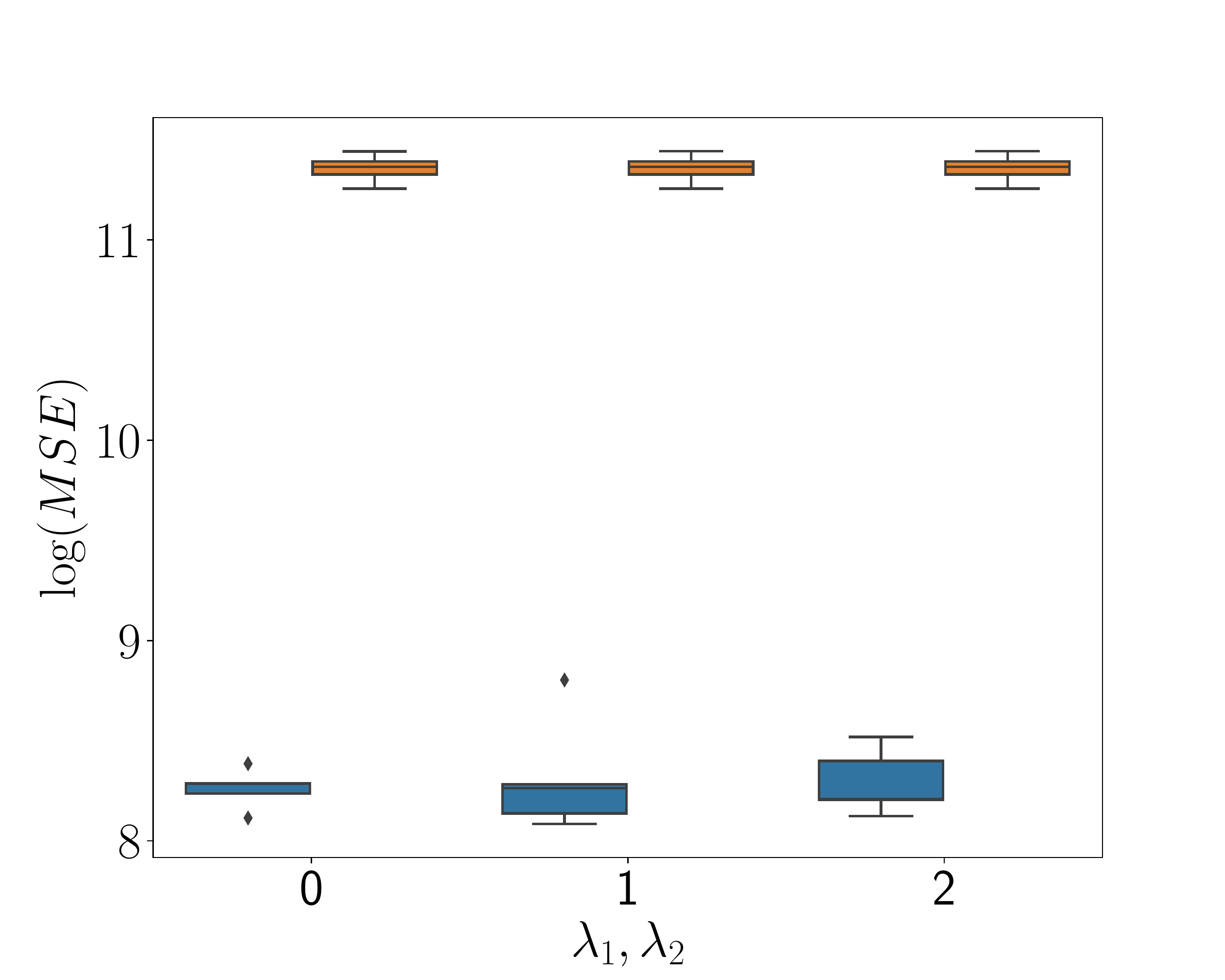} }
    \subfloat[\centering  ]{\includegraphics[width=0.33\textwidth]{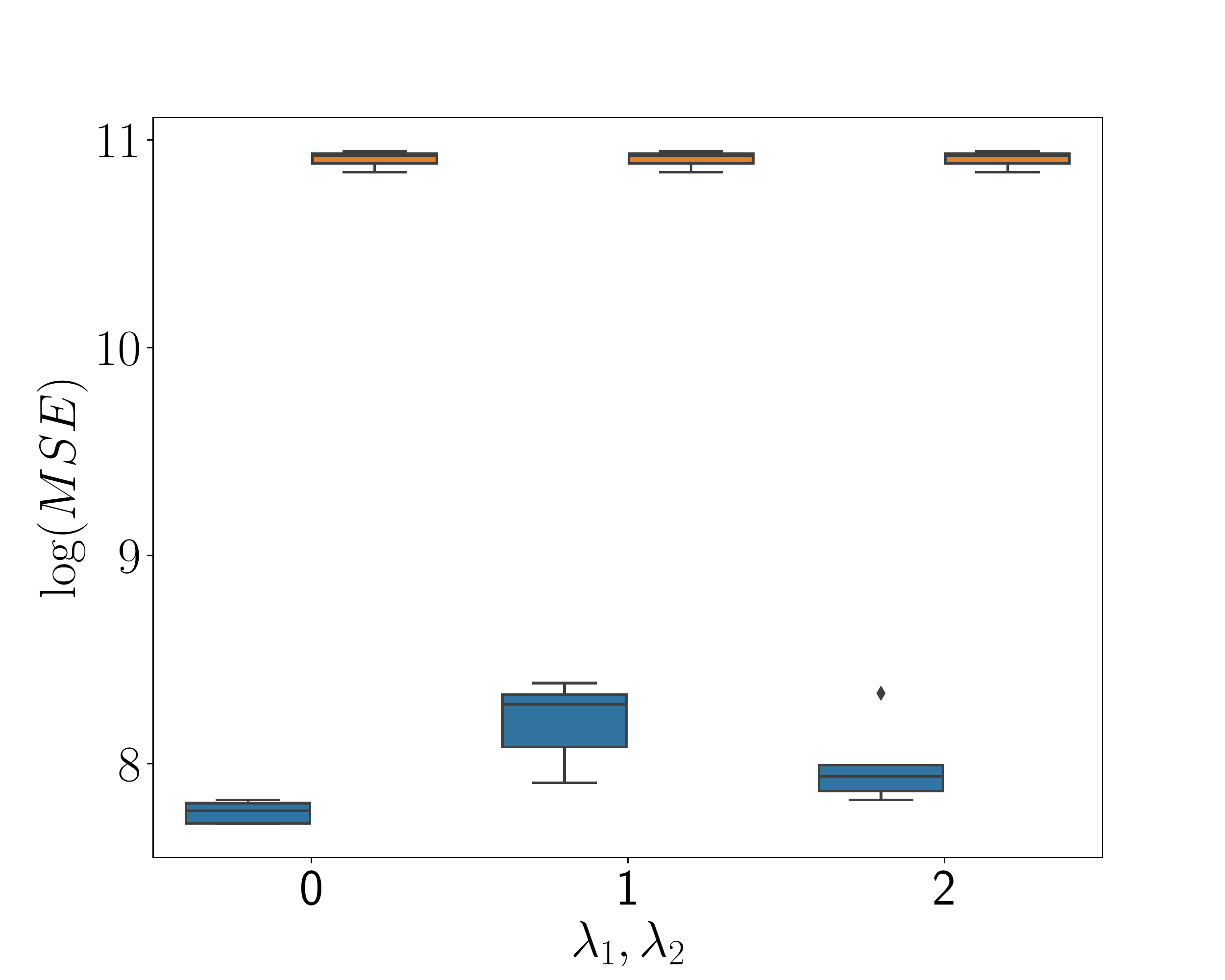} }
    \subfloat[\centering  ]{\includegraphics[width=0.33\textwidth]{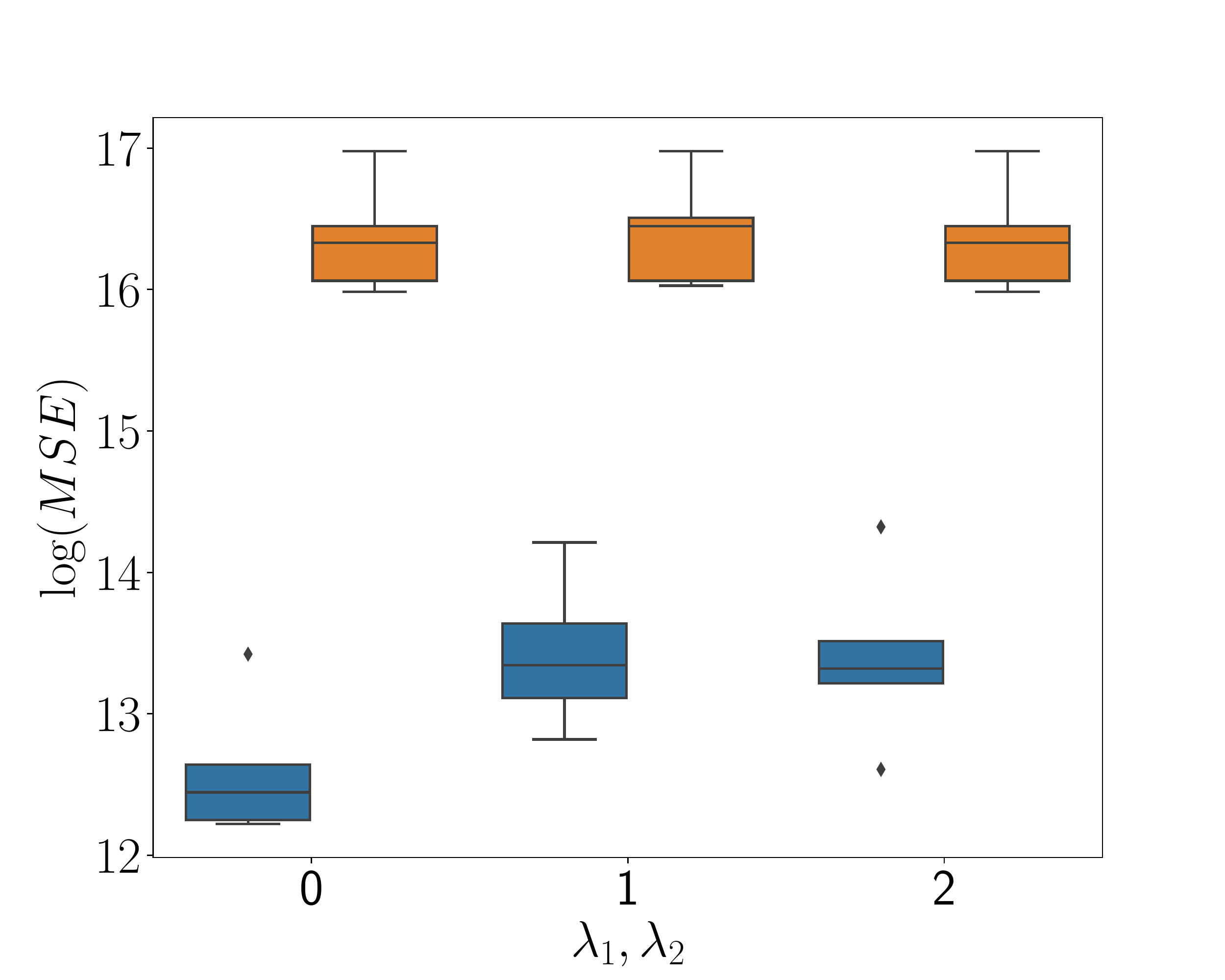} }
    \qquad
    \subfloat [\centering  ]{\includegraphics[width=0.33\textwidth]{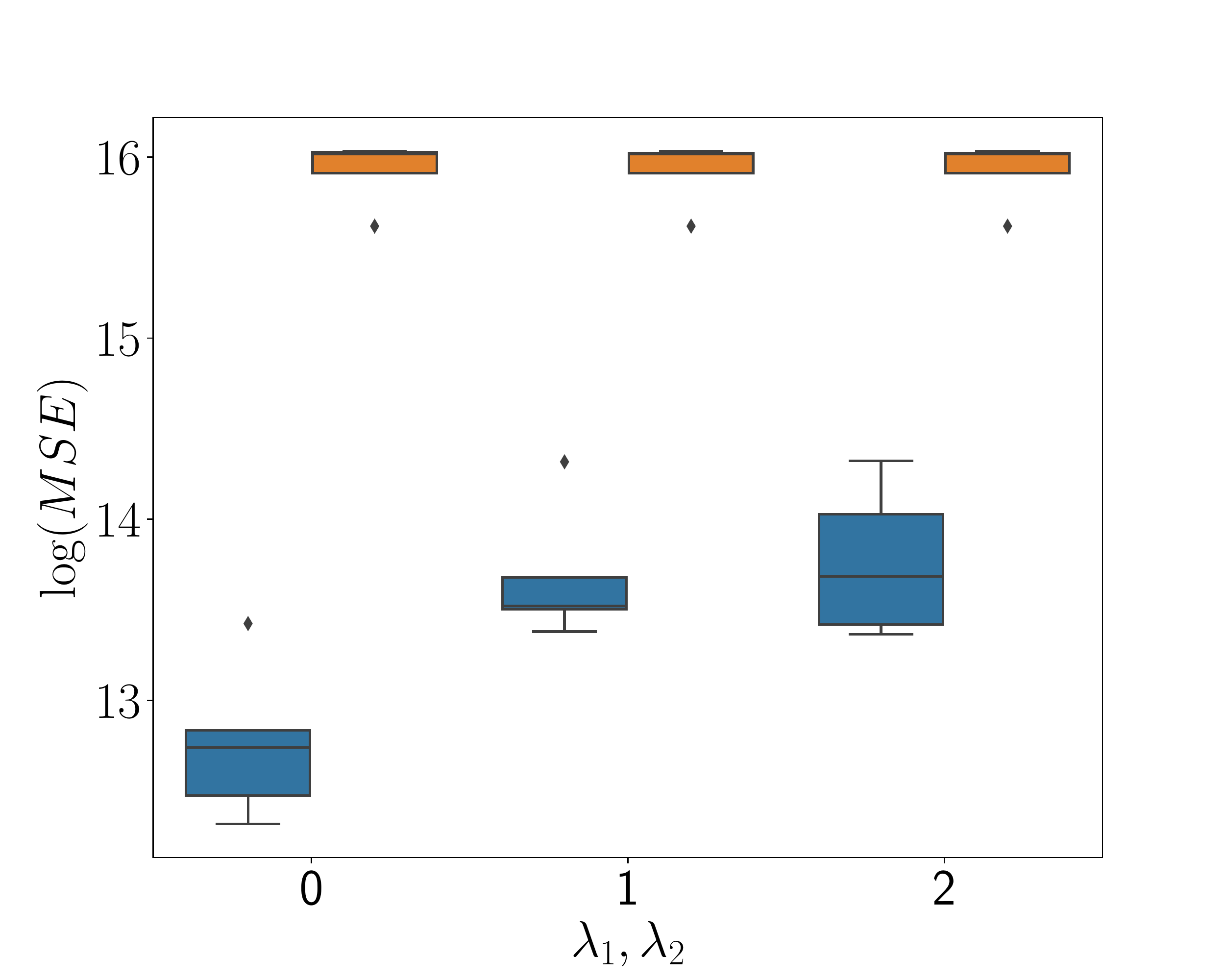} }
    \subfloat[\centering  ]{\includegraphics[width=0.33\textwidth]{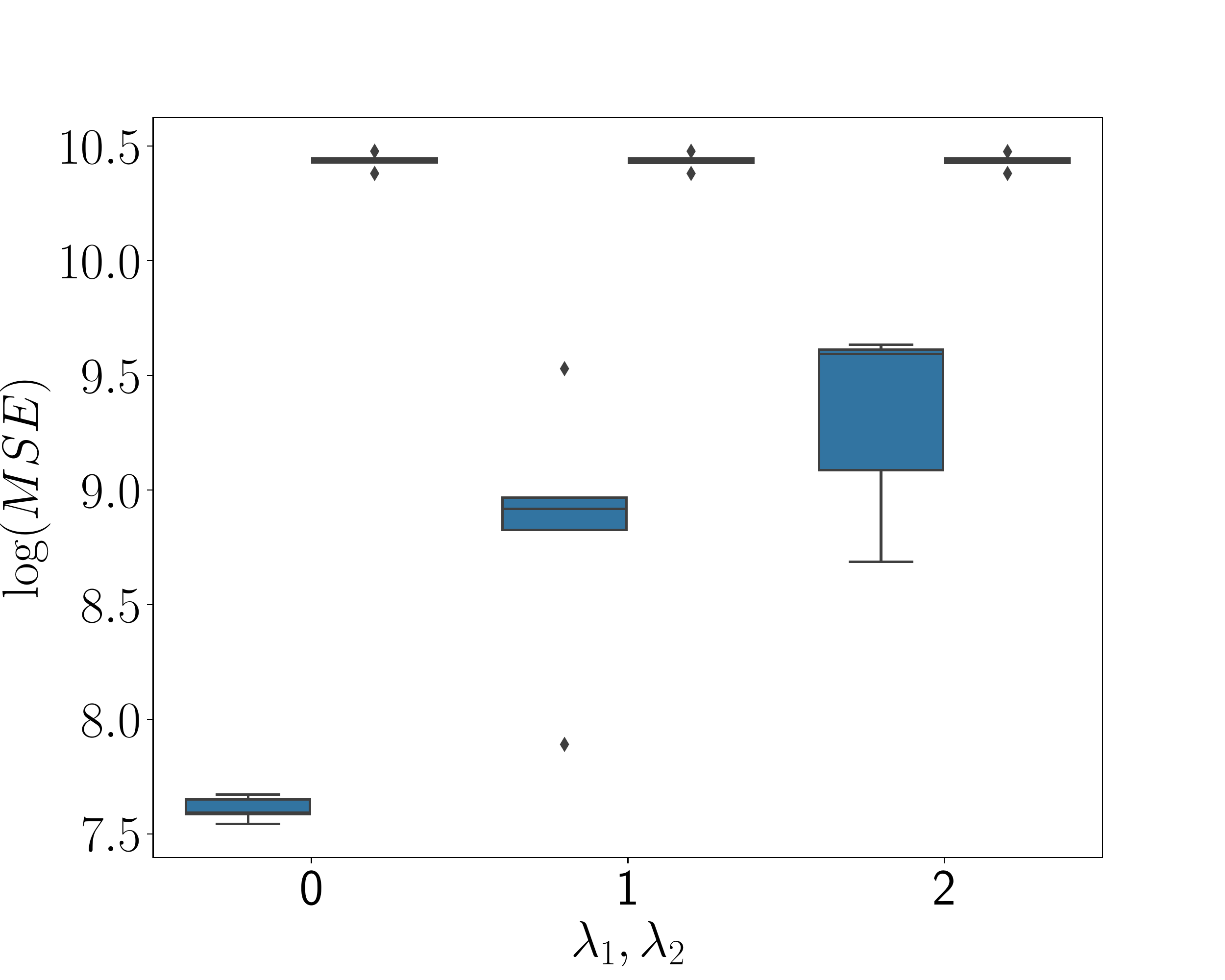} }
    \subfloat[\centering  ]{\includegraphics[width=0.33\textwidth]{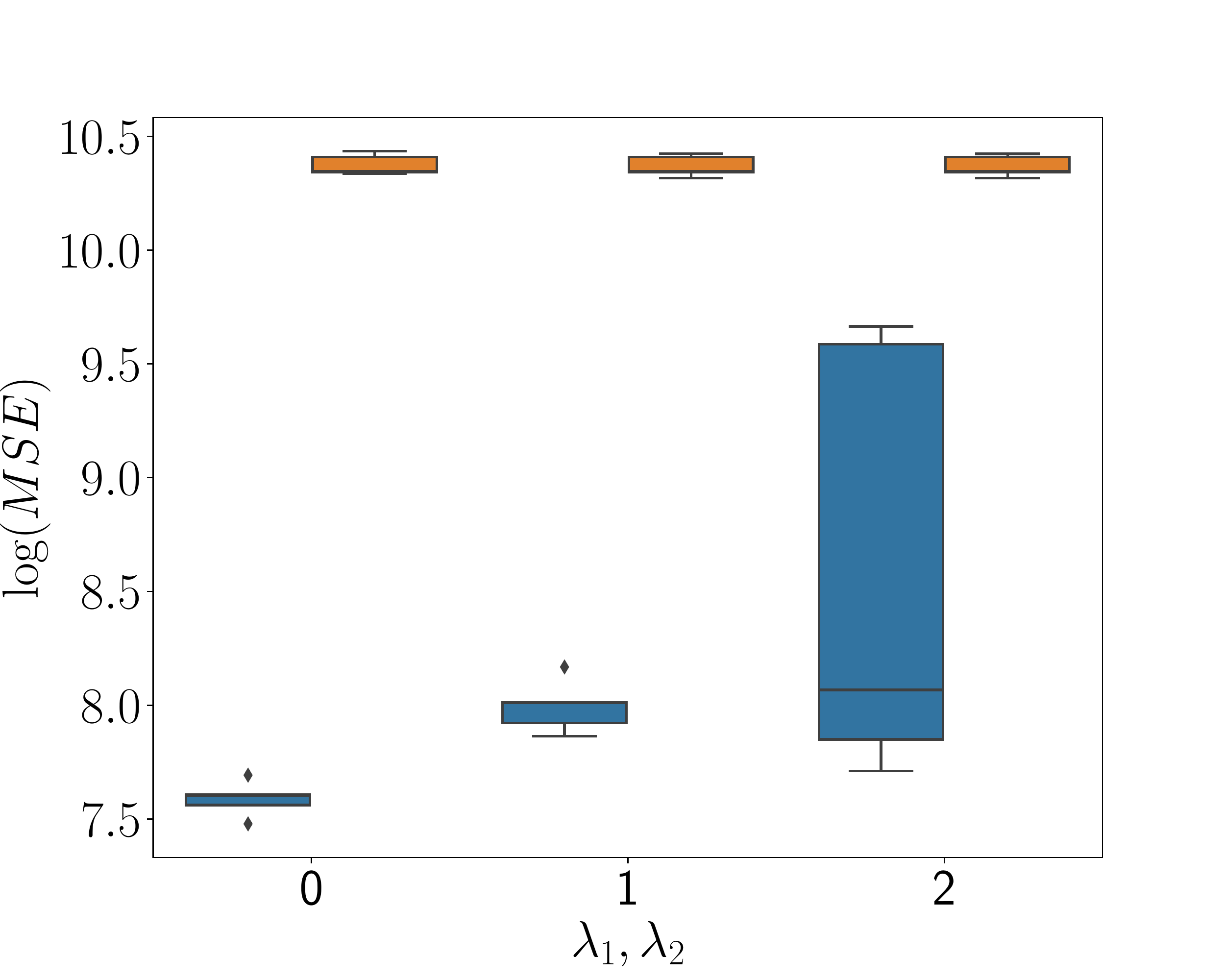} }
    \qquad
    \subfloat [\centering  ]{\includegraphics[width=0.33\textwidth]{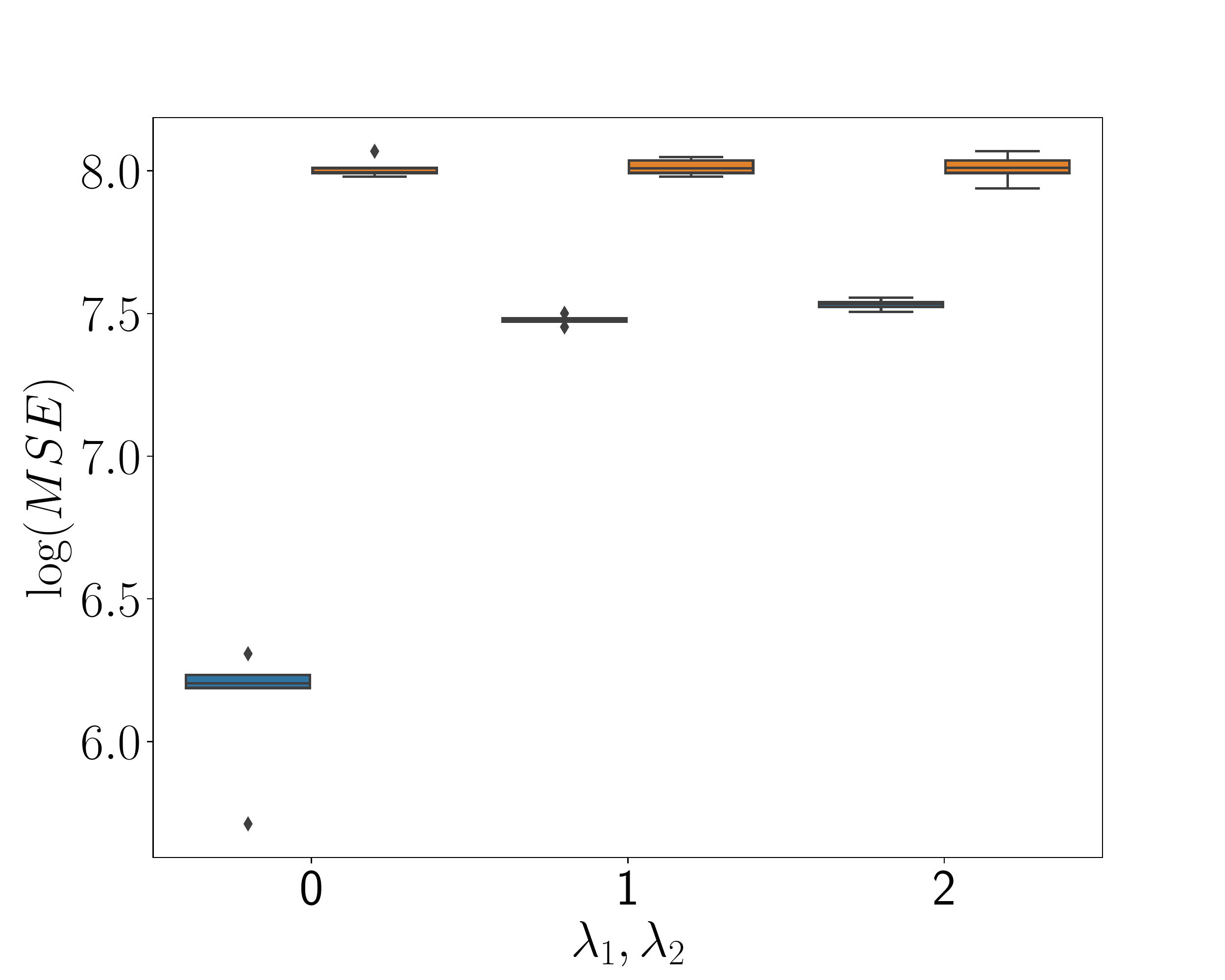} }
    \subfloat[\centering  ]{\includegraphics[width=0.33\textwidth]{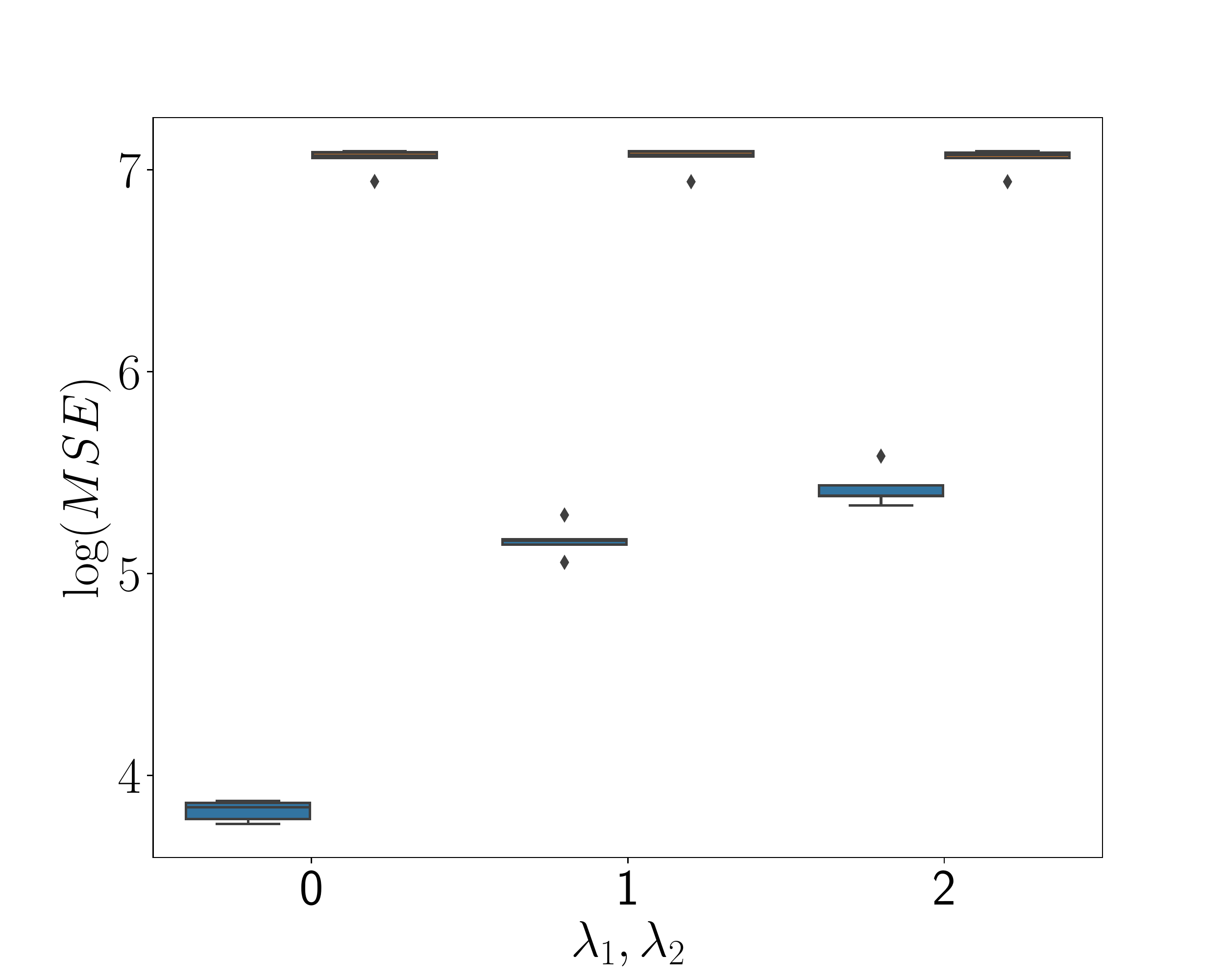} }
    \subfloat[\centering  ]{\includegraphics[width=0.33\textwidth]{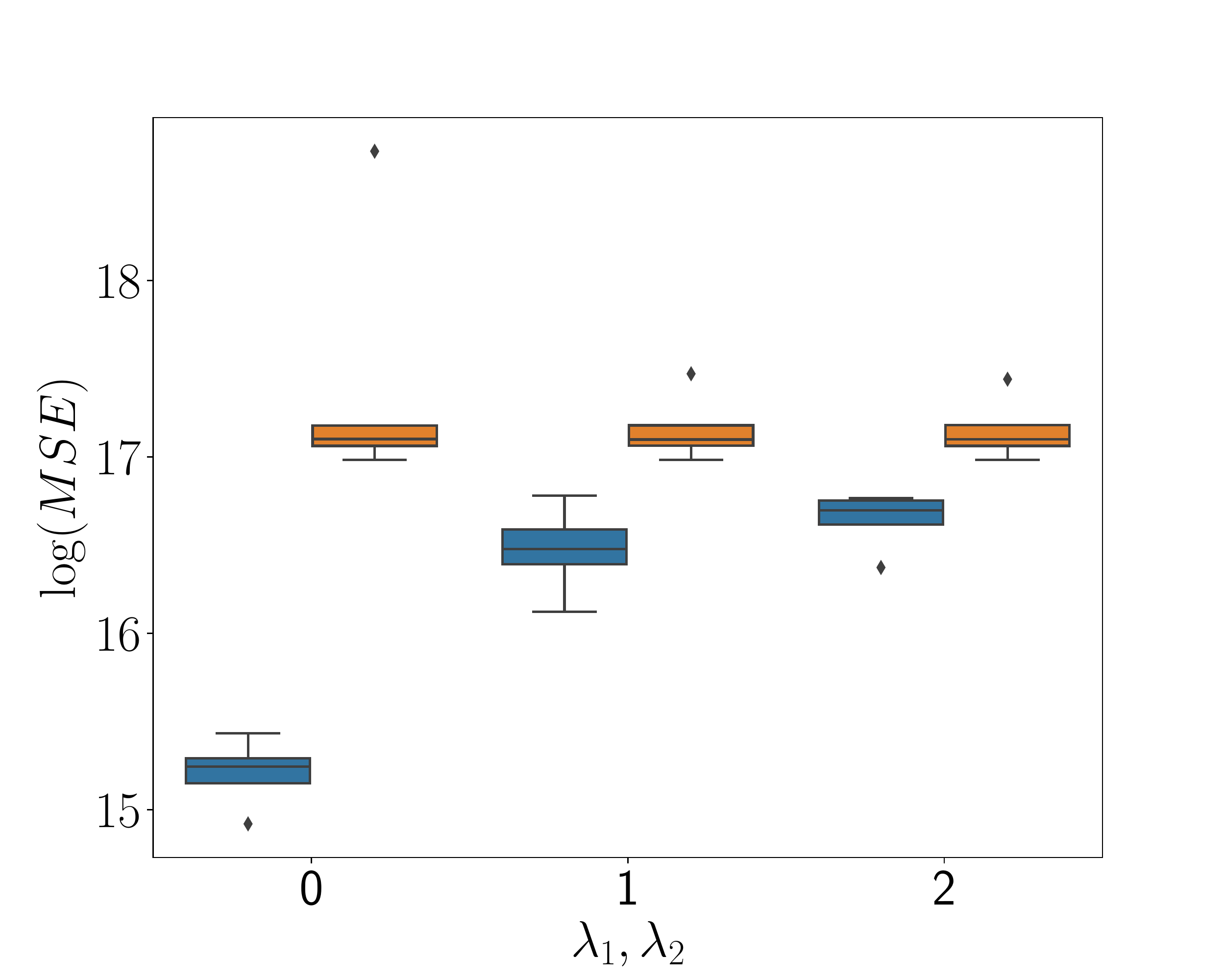} }
    \qquad
    \subfloat [\centering  ]{\includegraphics[width=0.33\textwidth]{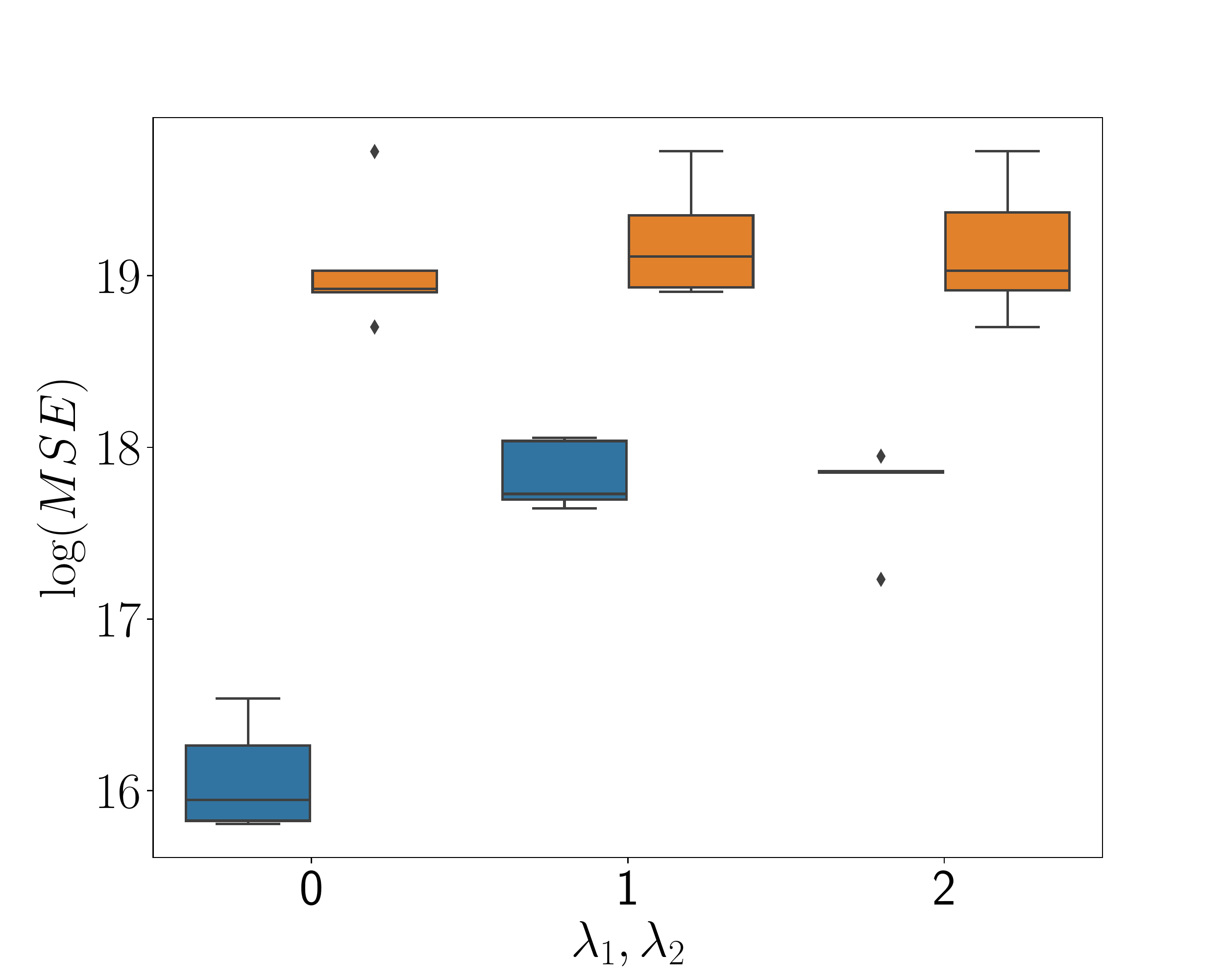} }
    \subfloat[\centering  ]{\includegraphics[width=0.33\textwidth]{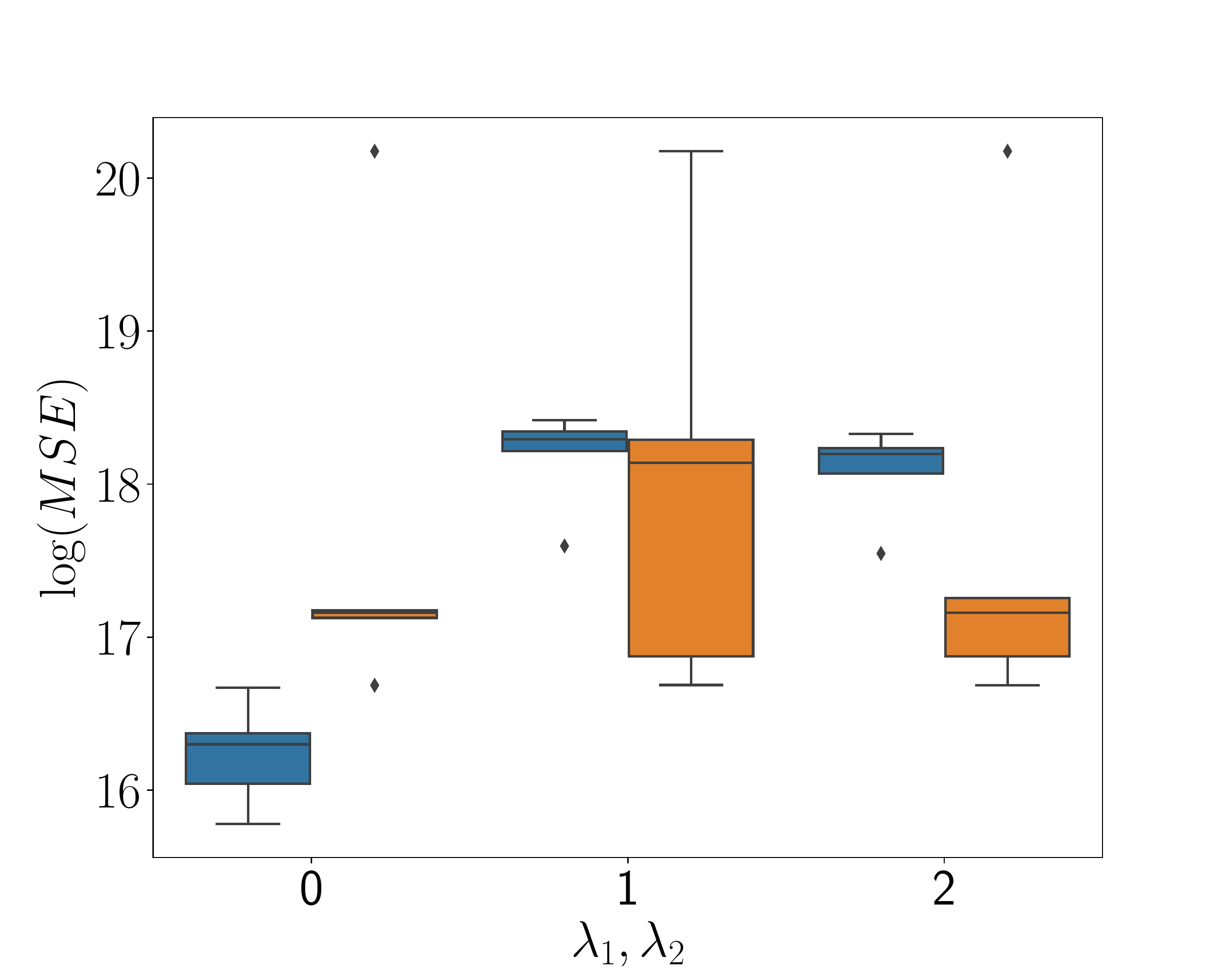} }
    \subfloat[\centering  ]{\includegraphics[width=0.33\textwidth]{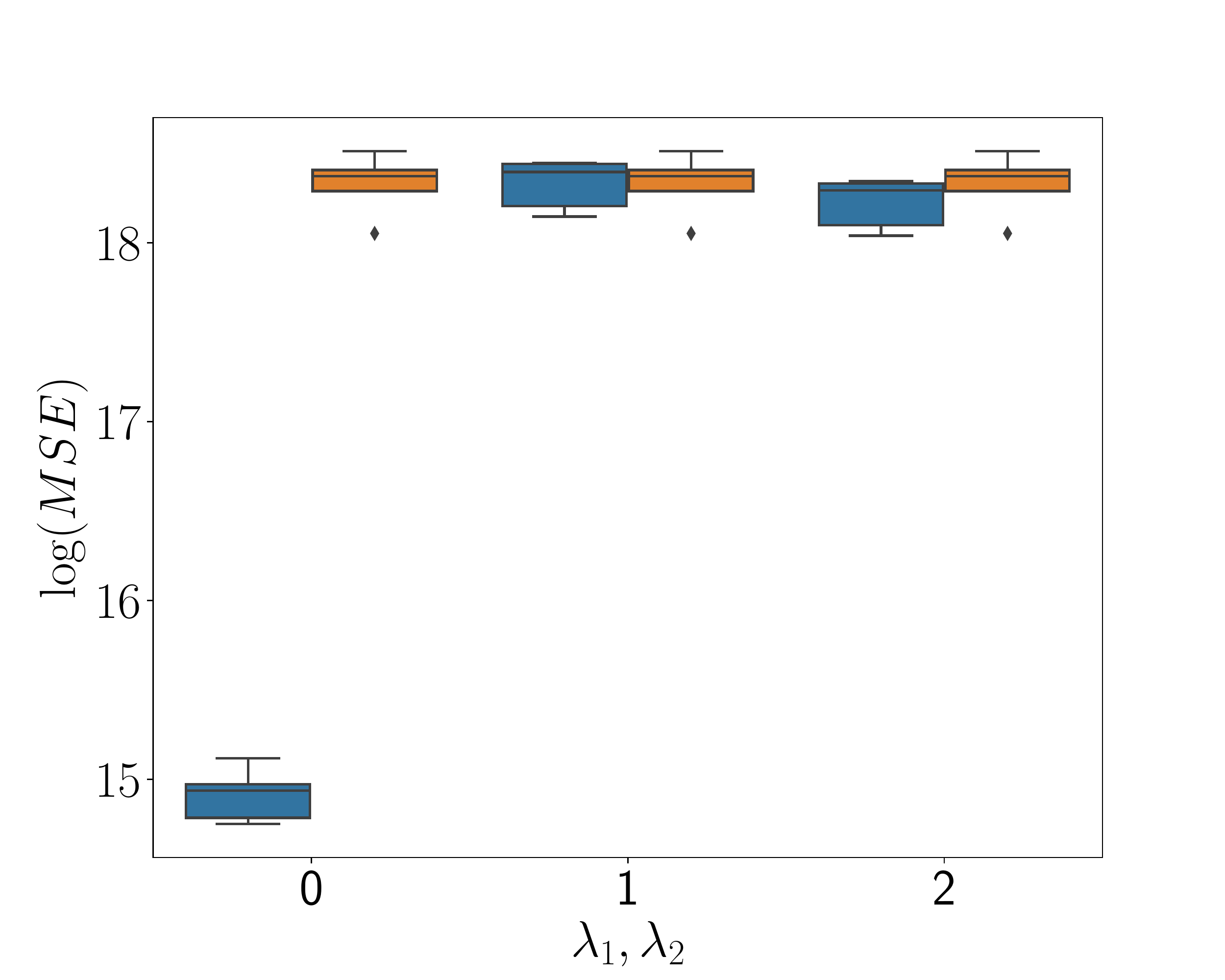} }
    \caption{Comparative performance in the unsupervised setting on simulated data. Blue: DNMF, orange: MU. (a-l) represent simulated data sets (1-12), respectively.}
    \label{fig: Unupervised DNMF-MU}%
\end{figure}

Next, we evaluate DNMF in the unsupervised case. In this case, the regularization parameters
are part of the objective function and cannot be learned by the model, hence we compare DNMF to MU 
under different regularization settings. As evident from the results in Figures~\ref{fig: Unupervised DNMF-MU} and~\ref{fig: Unupervised DNMF-MU, BRCA}, DNMF outperforms MU across a wide range of data sets and regularization values on both simulated and real data.


\begin{figure}
    \centering
    \includegraphics[width=0.80\textwidth]{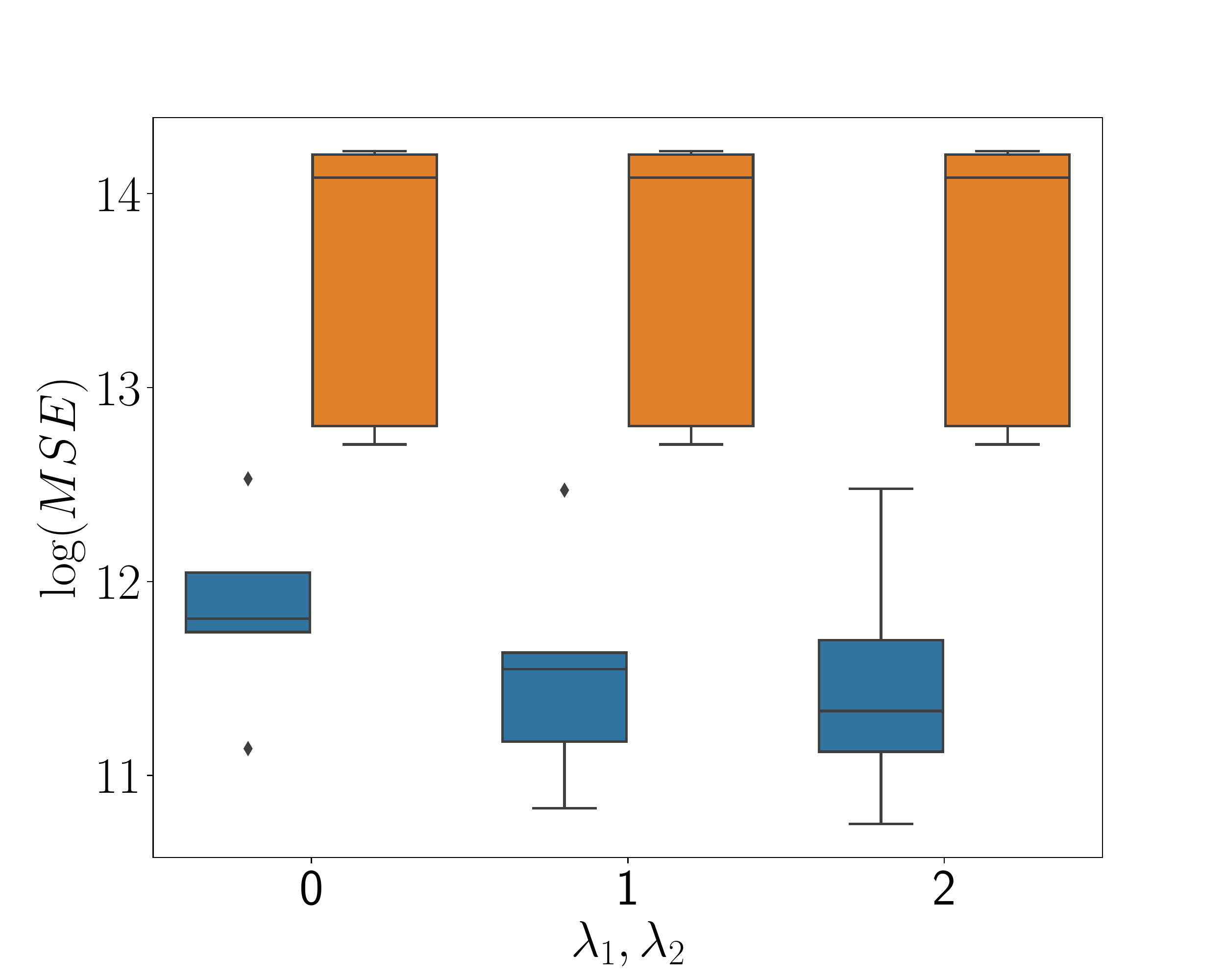}
    \caption{Comparative performance in the unsupervised setting on real data. Blue: DNMF, orange: MU.}
    \label{fig: Unupervised DNMF-MU, BRCA}%
\end{figure}

\begin{figure}
    \centering
    \subfloat [\centering $\lambda_1=\lambda_2=0$ ]{\includegraphics[width=0.33\textwidth]{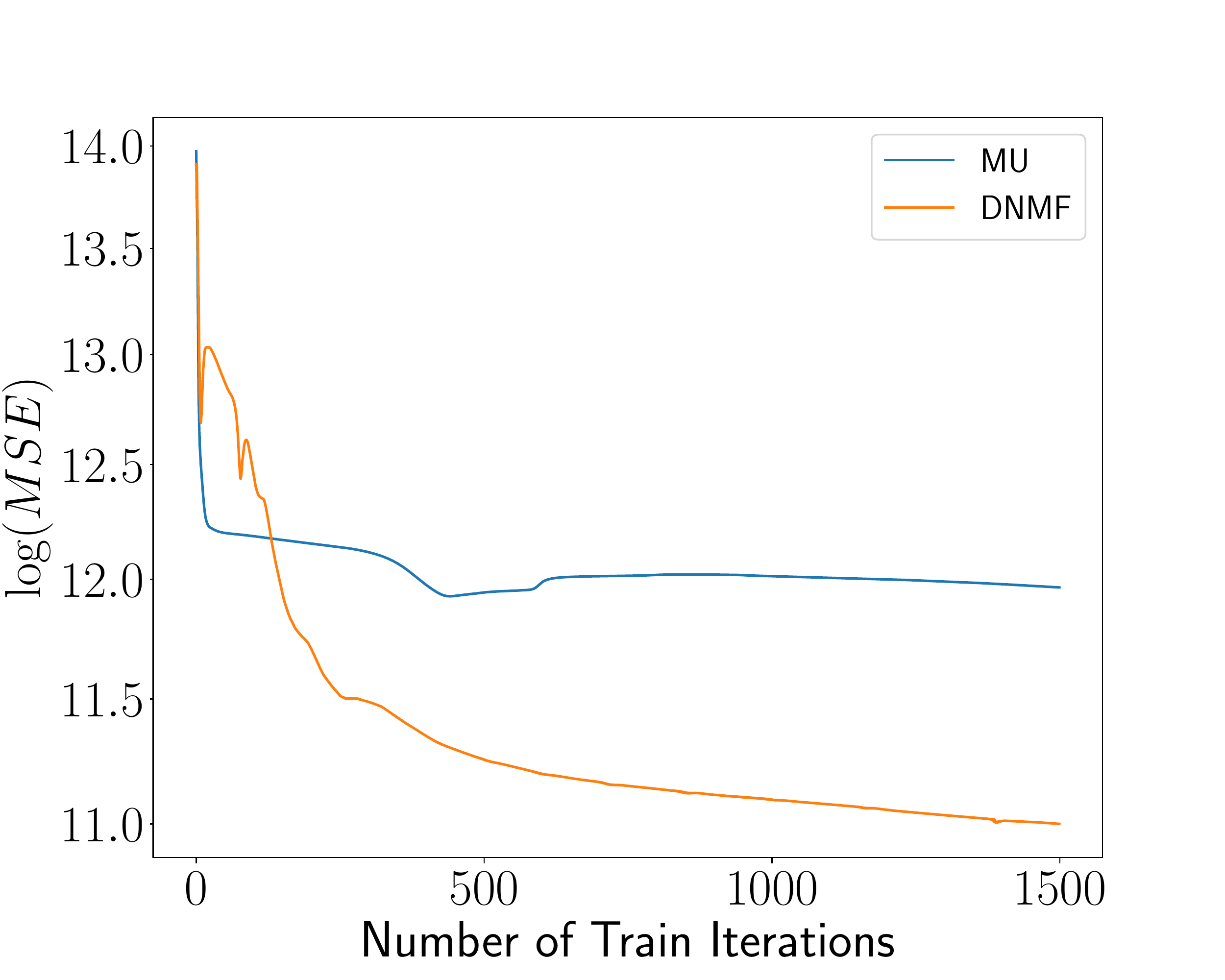} }
    \subfloat[\centering $\lambda_1=\lambda_2=1$ ]{\includegraphics[width=0.33\textwidth]{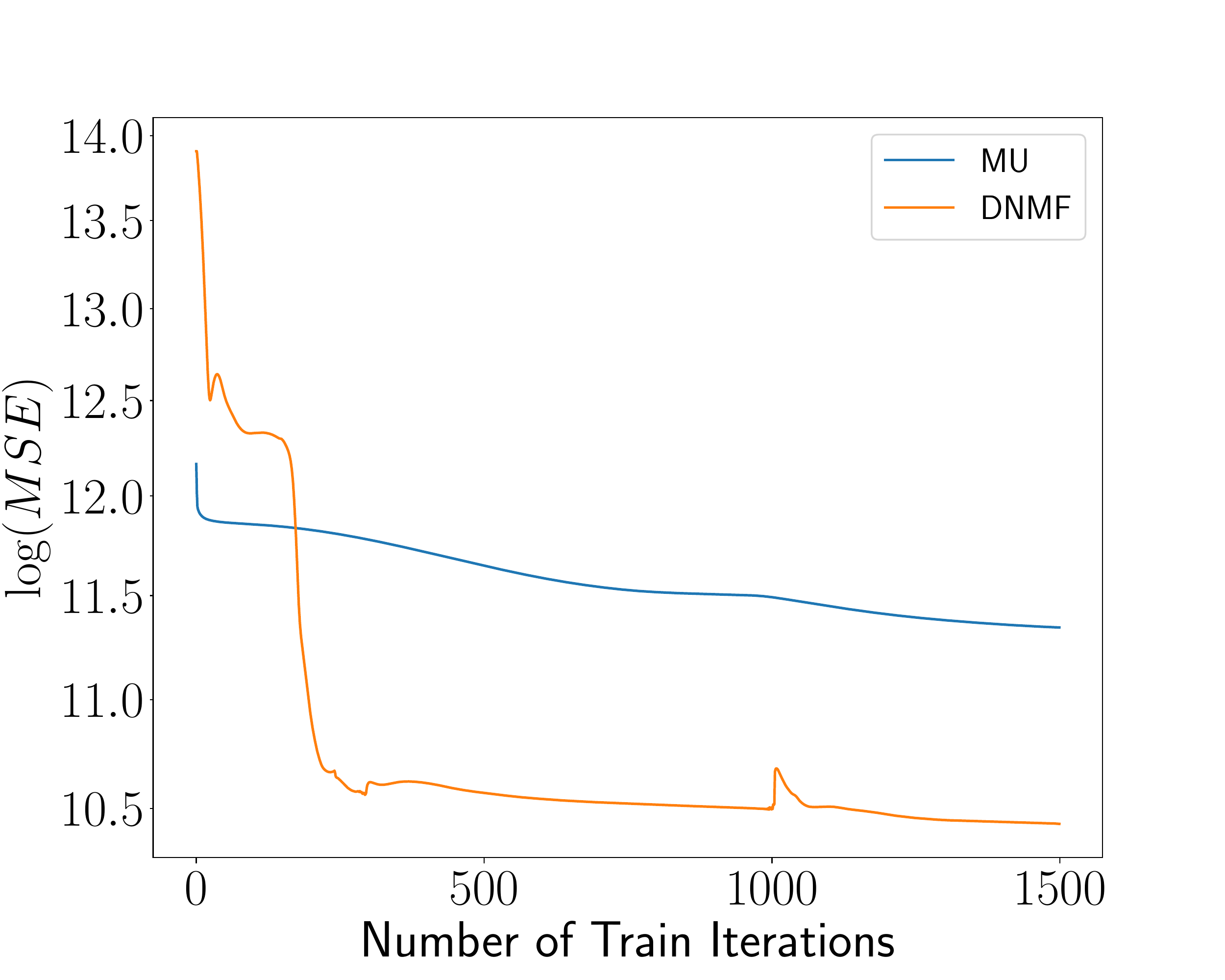} }
    \subfloat[\centering $\lambda_1=\lambda_2=2$ ]{\includegraphics[width=0.33\textwidth]{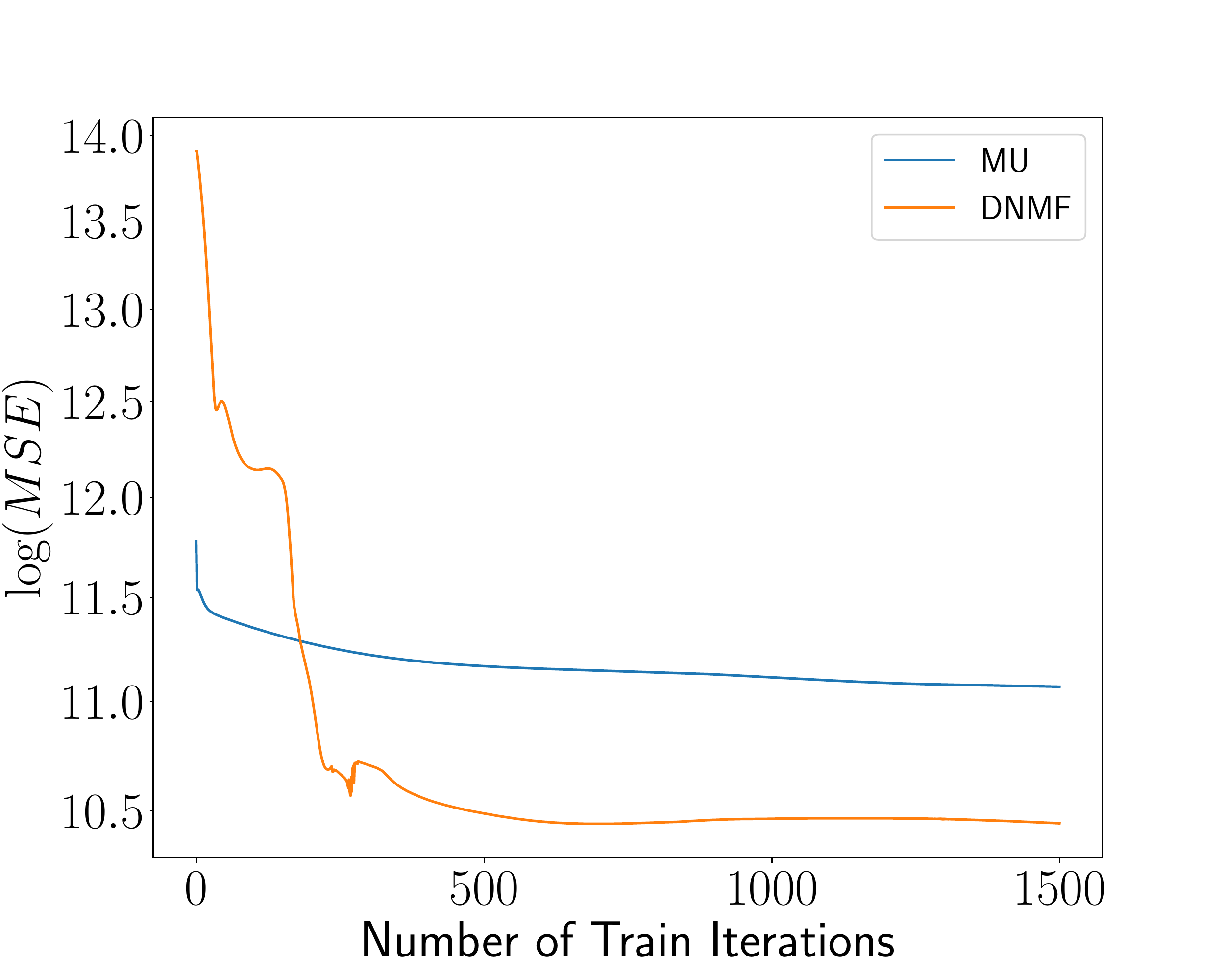} }
    \caption{Unsupervised reconstruction error during training on real data for MU and DNMF.}
    \label{fig: BRCA DNMF-MU Trainaning}%
\end{figure}

To get an intuition for the improved performance of DNMF compared to MU, we looked at the cost function
being optimized across algorithmic iterations when considering the real data set and multiple
regularization parameters. We observed that MU converges to a local minimum after a few iterations only,
hence we attempted different random initializations for it and report the best one. Nevertheless, DNMF remains the best performer under all settings (Figure~\ref{fig: BRCA DNMF-MU Trainaning}).

\section*{Conclusions}
We provided a detailed deep learning framework for non-negative matrix factorization that is applicable in both supervised and unsupervised settings.
The framework outperforms classical approaches to this problem and greatly improves the
reconstruction error of the factorizations across a wide range of data sets and regularization schemes.
We demonstrated the utility of our framework in analyzing mutation data from simulated and real data sets and expect it to greatly improve our ability to reconstruct mutational signatures and their exposures. 

For future work, we intend to explore different strategies for initializing the DNMF model and to select its regularization parameters in the unsupervised case.

\paragraph{Acknowledgments.}
We thank Itay Sason for his helpful feedback on the manuscript. 
RS was supported by the Israel Science Foundation (grant no. 2417/20), within the Israel Precision Medicine Partnership program.

\bibliographystyle{abbrv}
\bibliography{refs} 

\begin{thebibliography}{10}

\bibitem{Albright2006-wt}
R.~Albright, J.~Cox, D.~Duling, A.~N. Langville, and C.~Meyer.
\newblock Algorithms, initializations, and convergence for the nonnegative
  matrix factorization.
\newblock Technical report, Citeseer, 2006.

\bibitem{Alexandrov2019-fe}
L.~B. Alexandrov, J.~Kim, N.~J. Haradhvala, M.~N. Huang, A.~W.~T. Ng, Y.~Wu,
  A.~Boot, K.~R. Covington, D.~A. Gordenin, E.~N. Bergstrom, S.~M.
  Ashiqul~Islam, N.~Lopez-Bigas, L.~J. Klimczak, J.~R. McPherson,
  S.~Morganella, R.~Sabarinathan, D.~A. Wheeler, V.~Mustonen, {the PCAWG
  Mutational Signatures Working Group}, G.~Getz, S.~G. Rozen, and M.~R.
  Stratton.
\newblock The repertoire of mutational signatures in human cancer.
\newblock July 2019.

\bibitem{Boutsidis2008-vz}
C.~Boutsidis and E.~Gallopoulos.
\newblock {SVD} based initialization: A head start for nonnegative matrix
  factorization.
\newblock {\em Pattern Recognit.}, 41(4):1350--1362, Apr. 2008.

\bibitem{gillis2014and}
N.~Gillis.
\newblock The why and how of nonnegative matrix factorization.
\newblock {\em Regularization, optimization, kernels, and support vector
  machines}, 12(257):257--291, 2014.

\bibitem{Hershey2014-xw}
J.~R. Hershey, J.~L. Roux, and F.~Weninger.
\newblock Deep unfolding: {Model-Based} inspiration of novel deep
  architectures.
\newblock pages 1--27, 2014.

\bibitem{hoyer2002non}
P.~O. Hoyer.
\newblock Non-negative sparse coding.
\newblock In {\em Proceedings of the 12th IEEE Workshop on Neural Networks for
  Signal Processing}, pages 557--565. IEEE, 2002.

\bibitem{Kim2011-fj}
J.~Kim and H.~Park.
\newblock Fast nonnegative matrix factorization: An {Active-Set-Like} method
  and comparisons.
\newblock {\em SIAM J. Sci. Comput.}, 33(6):3261--3281, Jan. 2011.

\bibitem{lawson1995solving}
C.~L. Lawson and R.~J. Hanson.
\newblock {\em Solving least squares problems}.
\newblock SIAM, 1995.

\bibitem{Lee}
D.~D. Lee, M.~Hill, and H.~S. Seung.
\newblock Algorithms for non-negative matrix factorization.

\bibitem{Lin2007-tf}
C.-J. Lin.
\newblock Projected gradient methods for nonnegative matrix factorization.
\newblock {\em Neural Comput.}, 19(10):2756--2779, Oct. 2007.

\bibitem{monga2019algorithm}
V.~Monga, Y.~Li, and Y.~C. Eldar.
\newblock Algorithm unrolling: Interpretable, efficient deep learning for
  signal and image processing.
\newblock {\em arXiv preprint arXiv:1912.10557}, 2019.

\bibitem{vavasis2010complexity}
S.~A. Vavasis.
\newblock On the complexity of nonnegative matrix factorization.
\newblock {\em SIAM Journal on Optimization}, 20(3):1364--1377, 2010.

\bibitem{Wang2016-ay}
D.~Wang, J.-X. Liu, Y.-L. Gao, J.~Yu, C.-H. Zheng, and Y.~Xu.
\newblock An {NMF-L2,1-Norm} constraint method for characteristic gene
  selection.
\newblock {\em PLoS One}, 11(7):e0158494, July 2016.

\bibitem{Wisdom2017-ce}
S.~Wisdom, T.~Powers, J.~Pitton, and L.~Atlas.
\newblock Building recurrent networks by unfolding iterative thresholding for
  sequential sparse recovery.
\newblock {\em Proc. IEEE Int. Conf. Acoust. Speech Signal Process.}, pages
  4346--4350, 2017.

\end{thebibliography}

\appendix
\section{Regularized NMF}
\label{proof}

Consider the problem of finding an approximate non-negative factorization that is close to the original matrix $V$ and satisfies the sparseness constrains. We use the Frobenius norm as a measure of the distance between $V$ and $WH$, adding $L_1,L_2$ regularizations, arriving at the following cost function:
\begin{equation}
    C(W,H) = \frac{1}{2}\| V - WH \|^2_F + \lambda_1\| H \|_1 + \frac{1}{2}\lambda_2\| H \|_2^2.\ \label{eq:cost}
\end{equation}

\begin{theorem}
    The cost function is non-increasing under the update rules:
\[
H \gets H \odot \dfrac{W^TV}{W^TWH+\lambda_1+\lambda_2H} \hspace{0.1in};\hspace{0.1in} W \gets W \odot \dfrac{VH^T}{WHH^T}.
\]
\end{theorem}

\paragraph{Proof.}
We follow the proofs of~\cite{Lee,hoyer2002non} and focus on the update formula for $H$, 
considering one column $h$ at a time corresponding to a column $v$ of $V$.
Our goal is to minimize 
\[C(h)=\frac{1}{2}\|v-Wh\|^2_F+\lambda_1\|h\|_1+\frac{\lambda_2}{2}\|h\|_2^2.\]
As in these references, we define 
$G(h,h_l)$ to be an auxiliary function for $C(h)$ that satisfies
$G(h,h_l)\geq C(h), G(h,h) = C(h)$. At each iteration we update as follows:
\[
    h_{l+1} = \underset{h}{\operatorname{argmin}} \, G(h,h_l)
\]

We keep the original definition of the auxiliary function:
\[
    G(h,h_l) = C(h_l) + (h-h_l)^T\nabla C(h_l) + \dfrac{1}{2}(h-h_l)^T K(h_l)(h-h_l)
\]
where $K(h_l)$ is a diagonal matrix. However, we slightly change $K(h_l)$ to reflect the 
regularization: $K_{ab}(h_l):=K'_{ab}(h_l)+\lambda_2 = 
\delta_{ab}\frac{(W^TWh_l)_a+\lambda_1}{h_{la}}+\lambda_2$.
To show that $G(h,h_l) \geq C(h)$ we take a Taylor expansion of $C$:
\[
C(h) = C(h_l) + (h-h_l)^T\nabla C(h_l) + \dfrac{1}{2}(h-h_l)^T (W^TW+\lambda_2)(h-h_l).
\]
Thus, we need to show that $0 \leq (h-h_l)^T (K'(h_l)-W^TW)(h-h_l)$ which was shown in~\cite{hoyer2002non}.

It remains to compute the gradient of $G$ and equate it to zero:
\[
    \nabla_hG(h,h_l) = -\nabla C(h_l) + (h-h_l)K(h_l) = 0.
\]

This gives the update rule $h_{l+1} = h_l - K(h_l)^{-1}\nabla C(h_l)$ 
where $\nabla C(h_l) = -W^Tv+W^TWh_l+\lambda_2h_l+\lambda_1$.
Overall, we get:
\[
    h_{l+1} = h_l - \frac{h_l}{W^TWh_l+\lambda_1+\lambda_2h_l}(-W^Tv+W^TWh_l+\lambda_2h_l+\lambda_1) =
    h_l \odot \dfrac{W^Tv}{W^TWh_l+\lambda_1+\lambda_2h_l},
\]
completing the proof.
$\hfill\blacksquare$

\clearpage
\section{Simulated data}
\label{appendix:data}
\begin{table}[h!]
  \begin{center}
    \label{table1}
    \begin{filecontents*}{mut_description.csv}
number,Dataset,Samples,components
1,pancreas.sp,1000,11
2,pancreas.sa,1000,20
3,many.types.sp,2700,21
4,many.types.sa,2700,39
5,3.5.40.rcc.and.ovary.sp,1000,11
6,3.5.40.rcc.and.ovary.sa,1000,19
7,3.5.40.abst.sp,1000,3
8,3.5.40.abst.sa,1000,3
9,2.7a.7b.bladder.and.melanoma.sp,1000,11
10,2.7a.7b.bladder.and.melanoma.sa,1000,26
11,2.7a.7b.abst.sp,1000,3
12,2.7a.7b.abst.sa,1000,3
\end{filecontents*}
\pgfplotstabletypeset[
    col sep = comma,
    string replace*={_}{\textsubscript},
    every head row/.style={before row=\toprule,after row=\midrule},
    every last row/.style={after row=\bottomrule},
    display columns/0/.style={string type,column name={}},
    display columns/1/.style={string type,column name={Dataset}, column type = {l}},
    display columns/2/.style={string type,column name={\#Samples}},
    display columns/3/.style={string type,column name={\#Components}}
    ]{mut_description.csv}
    \caption{List of simulated data sets used in this study.}
    \label{table: data desc}
  \end{center}
\end{table}

\end{document}